\documentclass[aps,pra,twocolumn,showpacs,floatfix]{revtex4-1}

\usepackage{epsfig,amsmath,amssymb}
\usepackage{graphics} 
\usepackage{subfigure}
\usepackage{color}
\usepackage{multirow}
\usepackage{mathrsfs}
\usepackage{natbib}

\begin{document}

\title
{
A frustrated honeycomb-bilayer Heisenberg antiferromagnet: The spin-$\frac{1}{2}$ $J_{1}$--$J_{2}$--$J_{1}^{\perp}$ model
}

\author
{R. F. Bishop}
\email{raymond.bishop@manchester.ac.uk}
\author
{P. H. Y. Li}
\email{peggyhyli@gmail.com}
\affiliation
{School of Physics and Astronomy, Schuster Building, The University of Manchester, Manchester, M13 9PL, UK}


\begin{abstract}
 We use the coupled cluster method to study the zero-temperature
  quantum phase diagram of the spin-$\frac{1}{2}$
  $J_{1}$--$J_{2}$--$J_{1}^{\perp}$ model on the honeycomb bilayer
  lattice.  In each layer we include both nearest-neighbor and
  frustrating next-nearest-neighbor antiferromagnetic exchange
  couplings, of strength $J_{1}>0$ and
  $J_{2} \equiv \kappa J_{1} > 0$, respectively.  The two layers are
  coupled by an interlayer nearest-neighbor exchange, with coupling
  constant $J_{1}^{\perp} \equiv \delta J_{1}>0$.  We calculate
  directly in the infinite-lattice limit both the ground-state energy
  per spin and the N\'{e}el magnetic order parameter, as well as the
  triplet spin gap.  By implementing the method to very high orders of
  approximation we obtain an accurate estimate for the full boundary
  of the N\'{e}el phase in the $\kappa\delta$ plane.  For each value
  $\delta < \delta_{c}^{>}(0) \approx 1.70(5)$ we find an upper
  critical value $\kappa_{c}(\delta)$, such that N\'{e}el order is
  present for $\kappa < \kappa_{c}(\delta)$.  Conversely, for each
  value $\kappa < \kappa_{c}(0) \approx 0.19(1)$ we find an upper
  critical value $\delta_{c}^{>}(\kappa)$, such that N\'{e}el order
  persists for $0 < \delta < \delta_{c}^{>}(\kappa)$.  Most
  interestingly, for values of $\kappa$ in the range
  $\kappa_{c}(0) < \kappa < \kappa^{>} \approx 0.215(2)$ we find a
  reentrant behavior such that N\'{e}el order exists only in the range
  $\delta_{c}^{<}(\kappa) < \delta < \delta_{c}^{>}(\kappa)$, with $\delta_{c}^{<}(\kappa)>0$.  These latter
  upper and lower critical values coalesce when $\kappa = \kappa^{>}$,
  such that
  $\delta_{c}^{<}(\kappa^{>}) = \delta_{c}^{>}(\kappa^{>}) \approx
  0.25(5)$.
\end{abstract}


\maketitle

\section{INTRODUCTION}
\label{introd_sec}
Quantum magnets, comprising systems with a magnetic ion with spin
quantum number $s$ sitting on each of the $(N \to \infty)$ sites of an
extended regular periodic lattice, provide a rich playground for the
study of quantum many-body systems with exotic ground-state (GS)
phases that are entirely absent in their classical counterparts.
These latter systems here correspond to precisely the same
spin-lattice systems but in the limit where $s \to \infty$, such that
the spins become classical.  Of particular interest in this context is
the often subtle interplay that ensues between frustration and
quantum fluctuations.  Frustration may be either geometrically induced
[e.g., as on the two-dimensional (2D) triangular lattice] or
dynamically induced.  The latter is of special interest since it may
be tuned.

Thus, for example, one may vary the relative strengths of competing
interactions that are present in the model Hamiltonian under study,
and which tend in the classical $(s \to \infty)$ case to frustrate one
another in the sense that each interaction, in the absence of the
other, tends to promote a different form of magnetic long-range order
(LRO).  As is well known, such forms of frustration often tend to
promote the appearance of classical GS phases that are macroscopically
degenerate in energy.  For the quantum-mechanical counterparts (with a
finite value of $s$), the quantum fluctuations present, due to the fact that any
such stable classical GS phase is not an eigenstate of the quantum
Hamiltonian, then act particularly strongly among the degenerate set
of states.  Such a scenario then offers enhanced possibilities for the
complete suppression of quasiclassical magnetic LRO, with the
associated emergence of such exotic states as various valence-bond
crystalline (VBC) phases, multipolar or spin-nematic phases, and
quantum spin-liquid (QSL) phases.

In broad terms quantum fluctuations tend to be stronger, other things
being equal, for spin-lattice systems with lower values of each of the
parameters (i) spatial dimensionality $d$, (ii) spin quantum number
$s$, and (iii) lattice coordination number $z$.  The Mermin-Wagner
theorem \cite{Mermin:1966} excludes all forms of magnetic LRO in any
isotropic Heisenberg spin-lattice system with $d=1$, even at zero
temperature $(T=0)$, or with $d=2$, except precisely at $T=0$.  This
is due to the fact that for any such system it is impossible to break a
continuous symmetry.  For this reason 2D spin-lattice models at $T=0$
now occupy a special arena in which to study quantum phase
transitions.  From among the eleven 2D Archimedean lattices, the
honeycomb lattice is the simplest of the four that share the lowest
value, $z=3$, of the coordination number.  Of these four it is also
the most commonly occurring in real quasi-2D magnetic materials.
Frustrated spin-$\frac{1}{2}$ models on the 2D monolayer honeycomb
lattice have hence become intensively studied in recent years
\cite{Rastelli:1979_honey,Mattsson:1994_honey,Fouet:2001_honey,Mulder:2010_honey,Wang:2010_honey,Cabra:2011_honey,Ganesh:2011_honey_merge,*Ganesh:2011_honey_errata_merge,Clark:2011_honey,DJJF:2011_honeycomb,Reuther:2011_honey,Albuquerque:2011_honey,Mosadeq:2011_honey,Oitmaa:2011_honey,Mezzacapo:2012_honey,PHYLi:2012_honeycomb_J1neg,Bishop:2012_honeyJ1-J2,Bishop:2012_honey_circle-phase,Li:2012_honey_full,RFB:2013_hcomb_SDVBC,Ganesh:2013_honey_J1J2mod-XXX,Zhu:2013_honey_J1J2mod-XXZ,Zhang:2013_honey,Gong:2013_J1J2mod-XXX,Yu:2014_honey_J1J2mod}.
Much less work
\cite{Zhao:2012_honeycomb_s1,Gong:2015_honey_J1J2mod_s1,Bishop:2016_honey_grtSpins,Li:2016_honey_grtSpins,Li:2016_honeyJ1-J2_s1}
has been done on frustrated honeycomb-lattice monolayers comprising spins with
$s \geq 1$.  

The unfrustrated honeycomb-lattice Heisenberg antiferromagnet has
isotropic nearest-neighbor (NN) exchange interactions only, all with
equal strength $J_{1}>0$.  The quantum fluctuations, that are present
for all finite values of the spin quantum number $s$, act to reduce
partially the perfect N\'{e}el LRO present on the bipartite lattice in
the classical $(s \to \infty)$ limit.  Nevertheless, they do not
destroy it completely for any value of $s$
\cite{Li:2016_honey_grtSpins}, even for the lowest value
$s=\frac{1}{2}$.  Instead, the N\'{e}el sublattice magnetization $M$
is fractionally reduced in a monotonically increasing fashion as $s$
is reduced.  However, even for $s=\frac{1}{2}$, the N\'{e}el order
parameter $M$ takes a value equal to about 54\% of its classical
limiting value (see, e.g., Refs.\
\cite{Oitmaa:1992_honey,Low:2009_honey,Jiang:2012_honey,DJJFarnell:2014_archimedeanLatt,Bishop:2015_honey_low-E-param}).
Hence, to destroy the N\'{e}el LRO in the honeycomb monolayer requires
the addition of frustrating interactions.  Perhaps the simplest way to
do so is to include isotropic antiferromagnetic (AFM) Heisenberg
exchange interactions between next-nearest-neighbor (NNN) pairs of
spins, all with equal strength $J_{2}>0$, resulting in the so-called
$J_{1}$--$J_{2}$ model.  The corresponding $J_{1}$--$J_{2}$--$J_{3}$
model also includes isotropic Heisenberg exchange interactions between
next-next-nearest-neighbor (NNNN) pairs of spins with equal coupling
strength $J_{3}$.  The latter model has been studied particularly
along the line $J_{3}=J_{2}$ that includes the point
$J_{3}=J_{2}=\frac{1}{2}J_{1}$ of maximum classical frustration.  The
three classical phases that the model exhibits in the sector where
$J_{i}>0$ $(i=1,2,3)$ all meet at this triple point, at which the
classical GS phase is also macroscopically degenerate.

The spin-$\frac{1}{2}$ $J_{1}$--$J_{2}$--$J_{3}$ monolayer model, or
special cases of it (especially those with $J_{3}=0$ or
$J_{3}=J_{2}$), on the honeycomb lattice have been intensively
investigated by a variety of theoretical techniques in recent years
(see, e.g., Refs.\
\cite{Rastelli:1979_honey,Mattsson:1994_honey,Fouet:2001_honey,Mulder:2010_honey,Wang:2010_honey,Cabra:2011_honey,Ganesh:2011_honey_merge,*Ganesh:2011_honey_errata_merge,Clark:2011_honey,DJJF:2011_honeycomb,Reuther:2011_honey,Albuquerque:2011_honey,Mosadeq:2011_honey,Oitmaa:2011_honey,Mezzacapo:2012_honey,PHYLi:2012_honeycomb_J1neg,Bishop:2012_honeyJ1-J2,Bishop:2012_honey_circle-phase,Li:2012_honey_full,RFB:2013_hcomb_SDVBC,Ganesh:2013_honey_J1J2mod-XXX,Zhu:2013_honey_J1J2mod-XXZ,Zhang:2013_honey,Gong:2013_J1J2mod-XXX,Yu:2014_honey_J1J2mod}).
Although some open unsettled questions do still remain, nevertheless
by now there is a considerable degree of consensus about its overall
$T=0$ quantum phase diagram, as discussed more fully in Sec.\
\ref{model_sec}.  By contrast, much less attention has been devoted to
analogous bilayer models (see, e.g., Refs.\
\cite{Ganesh:2011_honey_merge,*Ganesh:2011_honey_errata_merge,Ganesh:2011_honey_bilayer_PRB84,Oitmaa:2012_honey_bilayer,Zhang:2014_honey_bilayer,Arlego:2014_honey_bilayer,Gomez:2016_honey_bilayer,Zhang:2016_honey_bilayer}),
even in the unfrustrated monolayer case where $J_{3}=J_{2}=0$, and
even though some experimentally studied quasi-2D materials with a
honeycomb-lattice structure are well modeled as honeycomb bilayers.  A
prime example is the bismuth oxynitrate layered material
Bi$_{3}$Mn$_{4}$O$_{12}$(NO$_{3}$)
\cite{Smirnova:2009:honey_spin_3half,Okubo:2010:honey_spin_3half}, in
which the spin-$\frac{3}{2}$ Mn$^{4+}$ ions are ordered in honeycomb
layers with no appreciable distortion.  More precisely, the crystal
field of the MnO$_{6}$ octahedral complexes, combined with a strong
Hund's rule coupling, leads in this compound to Heisenberg-like
moments on the Mn$^{4+}$ ions with an effective value $s=\frac{3}{2}$
for the spin quantum number.  Two layers of such Mn$^{4+}$ honeycomb
lattices are separated in Bi$_{3}$Mn$_{4}$O$_{12}$(NO$_{3}$) by bismuth
atoms.  The resulting bilayers are themselves well separated spatially
in this compound, and it seems therefore to be a good approximation to
take the bilayer honeycomb lattice as the relevant model in which to
describe its magnetic properties and behavior.  It has also been
suggested that a corresponding experimental realization of a
spin-$\frac{1}{2}$ Heisenberg model on the honeycomb bilayer could be
obtained by substituting V$^{4+}$ ions in place of the Mn$^{4+}$ ions
in Bi$_{3}$Mn$_{4}$O$_{12}$(NO$_{3}$).

Thus, both for potential experimental reasons and theoretically, since
the addition of an interlayer exchange coupling is another way of
destroying the N\'{e}el order in honeycomb-lattice monolayers with NN
interactions only, it is undoubtedly of interest to study Heisenberg
models on honeycomb-lattice bilayers, in which competing frustration
on top of the NN intralayer AFM exchange couplings $J_{1}$ is present
due both to NNN intralayer AFM exchange couplings $J_{2}$ and to NN
interlayer AFM exchange couplings $J_{1}^{\perp}$.  We shall study the
resulting $J_{1}$--$J_{2}$--$J_{1}^{\perp}$ model here for the case
$s=\frac{1}{2}$.  Since the coupled cluster method (CCM) has already
been applied to the $s=\frac{1}{2}$ $J_{1}$--$J_{2}$--$J_{3}$
Heisenberg model on the honeycomb monolayer (or to special cases of
it, e.g., with $J_{3}=J_{2}$ or $J_{3}=0$) with great success
\cite{DJJF:2011_honeycomb,PHYLi:2012_honeycomb_J1neg,Bishop:2012_honeyJ1-J2,Bishop:2012_honey_circle-phase,Li:2012_honey_full,RFB:2013_hcomb_SDVBC},
and where it has been shown to give a good description of the $T=0$
phase diagram of the model, including estimates for its quantum
critical points (QCPs) which are among the best currently available,
we also use the method here.

We note from the outset that the spin-$\frac{1}{2}$
$J_{1}$--$J_{2}$--$J_{1}^{\perp}$ model on the bilayer honeycomb
lattice has also been studied recently
\cite{Zhang:2014_honey_bilayer,Arlego:2014_honey_bilayer} using a
Schwinger-boson parametrization of the spin operators followed by a
mean-field decoupling.  The results obtained for the model by this
Schwinger-boson mean field theory approach were also augmented by, and
compared with, corresponding results from both the exact
diagonalization of a small (24-site) cluster and a dimer-series
expansion calculation carried out to (the relatively low) fourth-order
for the triplet spin energy gap
\cite{Zhang:2014_honey_bilayer,Arlego:2014_honey_bilayer}.  It is thus
of particular interest to compare our own results from a potentially
much more accurate high-order implementation of the CCM with these
earlier results.  As a foretaste we note that while the two sets of
results show many qualitative similarities (e.g., for the overall
shape of the N\'{e}el phase boundary), significant quantitative
differences are also found, as discussed more fully in Sec.\
\ref{discuss_summary_sec}.

In Sec.\ \ref{model_sec} we first describe the model itself, including
a description of the salient features of the limiting case,
$J_{1}^{\perp}=0$, of the monolayer model.  We then briefly describe
the main elements of the CCM in Sec.\ \ref{ccm_sec}, before presenting
our results in Sec.\ \ref{results_sec}.  We conclude with a discussion
and summary in Sec.\ \ref{discuss_summary_sec}.

\section{THE MODEL}
\label{model_sec}
\begin{figure}[t]
\begin{center}
\mbox{
\subfigure[]{\includegraphics[width=4.0cm]{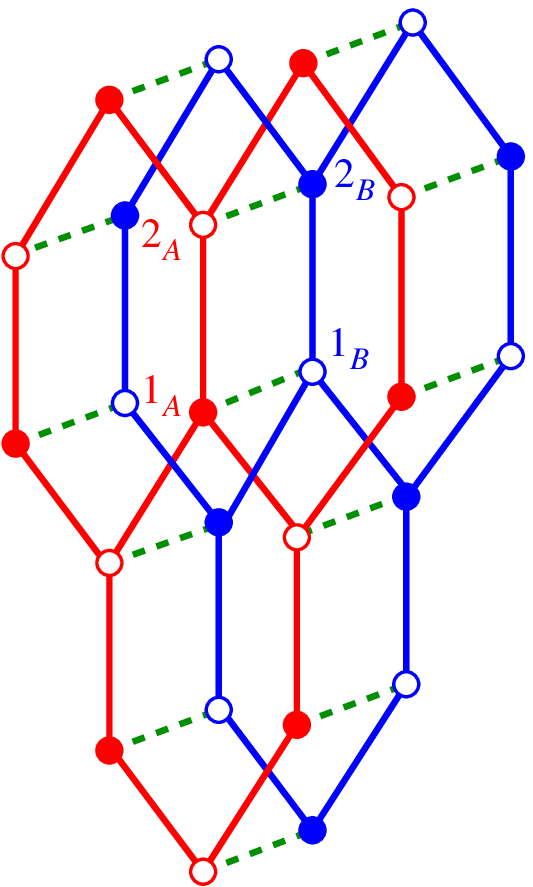}}
\quad \subfigure[]{\includegraphics[width=3.0cm]{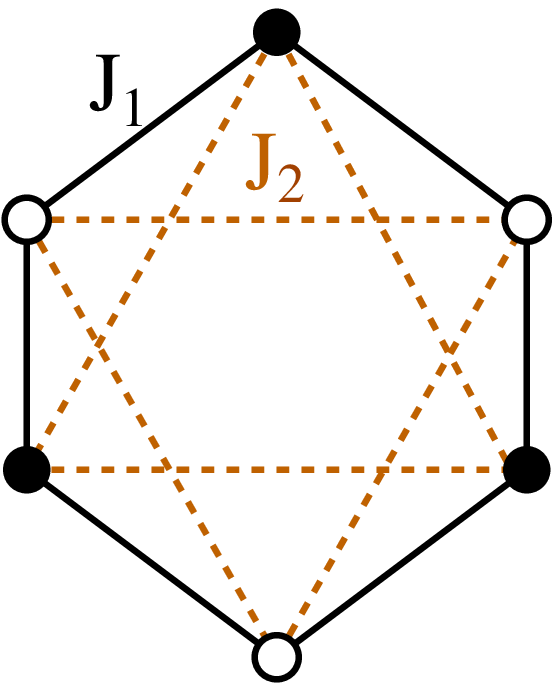}}
  }
  \caption{The $J_{1}$--$J_{2}$--$J_{1}^{\perp}$ model on the honeycomb bilayer
    lattice, showing (a) the two layers $A$ (red) and $B$ (blue), the nearest-neighbor bonds ($J_{1} = $ -----; $J_{1}^{\perp} = $ - - -), and the four sites ($1_{A}, 2_{A}, 1_{B}, 2_{B}$) of the unit cell; and (b) the intralayer bonds $J_{1} = $ -----; $J_{2} = $ - - -) on each layer.  Sites on the two monolayer triangular sublattices are shown by filled and empty circles respectively.}
\label{model_pattern}
\end{center}
\end{figure}

The Hamiltonian of the $J_{1}$--$J_{2}$--$J_{1}^{\perp}$ model on the
honeycomb bilayer lattice is given by
\begin{eqnarray}
H=&&J_{1}\sum_{{\langle i,j \rangle},\alpha} \mathbf{s}_{i,\alpha}\cdot\mathbf{s}_{j,\alpha} + 
J_{2}\sum_{{\langle\langle i,k \rangle\rangle},\alpha} \mathbf{s}_{i,\alpha}\cdot\mathbf{s}_{k,\alpha}  \nonumber \\
&&+ J_{1}^{\perp}\sum_{i} \mathbf{s}_{i,A}\cdot\mathbf{s}_{i,B}\,,
\label{H_eq}
\end{eqnarray}
where the index $\alpha=A,B$ denotes the two layers, and where each
site $i$ of the honeycomb lattice on each of the two layers carries a
spin-$s$ particle described by the SU(2) spin operator
${\bf
  s}_{i,\alpha}\equiv(s^{x}_{i,\alpha},s^{y}_{i,\alpha},s^{z}_{i,\alpha})$,
with ${\bf s}^{2}_{i,\alpha} = s(s+1)$.  For the case considered here,
$s=\frac{1}{2}$.  The sums over $\langle i,j \rangle$ and
$\langle \langle i,k \rangle \rangle$ in Eq.\ (\ref{H_eq}) run over
all intralayer NN and NNN bonds, respectively, on each
honeycomb-lattice monolayer, counting each pairs of spins once and
once only in each of the two sums.  The last sum in Eq.\ (\ref{H_eq})
over the index $i$ thus includes all NN interlayer pairs.  We shall be
interested here in the case when all three bonds are AFM in nature.
With no loss of generality we may henceforth put $J_{1} \equiv 1$ to
set the overall energy scale.  The (monolayer) honeycomb lattice is
non-Bravais with two sites per unit cell.  It comprises two
interlacing triangular Bravais sublattices, shown by filled and empty
circles respectively in Fig.\ \ref{model_pattern}.  The honeycomb
bilayer lattice thus has four sites per unit cell, also as shown in
Fig.\ \ref{model_pattern}.

In order to set the scene for our later discussion of the AFM
honeycomb bilayer lattice, we first comment on the situation for the
corresponding monolayer lattice (i.e., when $J_{1}^{\perp}=0$).  In
this case the classical ($s \to \infty$) $J_{1}$--$J_{2}$ model on the
honeycomb lattice has N\'{e}el order (i.e., where all of the spins on
the lattice sites denoted by filled circles in Fig.\
\ref{model_pattern} point in a given, arbitrary, direction, and those
on the sites denoted by empty circles point in the opposite direction)
for all values $0 \leq J_{2}/J_{1} \leq \frac{1}{6}$ of the intralayer
frustration parameter.  For values $J_{2}/J_{1} > \frac{1}{6}$ the
classical monolayer spins acquire spiral order
\cite{Rastelli:1979_honey,Fouet:2001_honey}.  In fact the spiral wave
vector of the corresponding GS phase can point in an arbitrary
direction and there exists an infinite one-parameter family of states,
all of which are degenerate in energy.  It has been shown that at the
level of lowest-order spin-wave theory, in which the leading
correction in the parameter $1/s$ is considered, this accidental
degeneracy is then lifted by spin-wave fluctuations, so that particular
spiral wave vectors are favored, thereby leading to so-called spiral
order by disorder \cite{Mulder:2010_honey}.

In more detail, in the region
$\frac{1}{6} \leq J_{2}/J_{1} \leq \frac{1}{2}$ two distinct spiral
phases, called the spiral-I and spiral-II phases, coexist, while for
$J_{2}/J_{1} > \frac{1}{2}$ the stable GS phase comprises only the
spiral-I phase \cite{Rastelli:1979_honey,Fouet:2001_honey}.  In fact,
for the larger classical $J_{1}$--$J_{2}$--$J_{3}$ model on the
honeycomb monolayer, the two points
$(J_{2}/J_{1},J_{3}/J_{1}) = (\frac{1}{6},0), (\frac{1}{2},0)$ are
tricritical points \cite{Fouet:2001_honey}.  At the former point the
two spiral phases meet the N\'{e}el phase, while at the latter point
they meet another collinear AFM phase, called the N\'{e}el-II phase,
described below.  More generally, in the classical limit
$(s \to \infty)$, there exists an infinitely degenerate family of
noncoplanar states, all of whom are degenerate in energy with the
N\'{e}el-II states, for every pair of values of the exchange coupling
constants $J_{2}/J_{1}$ and $J_{3}/J_{1}$ for which the N\'{e}el-II
state exists as a stable GS phase.  It has been shown
\cite{Fouet:2001_honey} that both thermal and quantum fluctuations
then favor the collinear N\'{e}el-II AFM phase, thereby lifting the
otherwise accidental degeneracy at the fully classical level.

Just like the N\'{e}el phase, the N\'{e}el-II phase also consists of
sets of parallel zigzag (or sawtooth) AFM chains along one of the
three equivalent honeycomb directions.  However, whereas NN spins on
adjacent chains are also antiparallel for the N\'{e}el state, for the
N\'{e}el-II state they are parallel.  Hence, equivalently, the
N\'{e}el and N\'{e}el-II states have, respectively, all 3 and 2 NN
pairs of spins aligned antiparallel to one another.  There are thus
three equivalent N\'{e}el-II states, one each corresponding to one of
the three fundamental honeycomb-lattice directions.  Each has a
four-sublattice (i.e., a 4-site unit cell) structure, and each clearly
breaks the lattice rotational symmetry, unlike the N\'{e}el state.

For the corresponding quantum versions of the $J_{1}$--$J_{2}$
honeycomb-lattice model (i.e., with finite values of the spin quantum
number $s$), it is to be expected that quantum fluctuations will act
to destroy the spiral order that is present in the classical
$(s \to \infty)$ case for $J_{2}/J_{1} \geq \frac{1}{6}$.  For the
case $s=\frac{1}{2}$ there is wide consensus via a variety of
theoretical calculational schemes that this indeed is the case.  It
has been shown by the CCM technique used here, for example, that for
the spin-$\frac{1}{2}$ case spiral order is absent over the entire
range $0 \leq J_{2}/J_{1} \leq 1$
\cite{Bishop:2012_honeyJ1-J2,RFB:2013_hcomb_SDVBC}.  It is also to be
expected, {\it a  priori}, that since quantum fluctuations generally
tend to favor collinear order, the critical value of $J_{2}/J_{1}$
above which N\'{e}el order melts in the spin-$\frac{1}{2}$
$J_{1}$--$J_{2}$ model on the honeycomb lattice will be {\it greater}
than the classical value of $\frac{1}{6}$.  A large variety of different
calculations now concurs with this expectation, giving a corresponding
critical value for the $s=\frac{1}{2}$ case of about 0.2 (see, e.g.,
Refs.\
\cite{Albuquerque:2011_honey,Mosadeq:2011_honey,Mezzacapo:2012_honey,Bishop:2012_honeyJ1-J2,RFB:2013_hcomb_SDVBC,Ganesh:2013_honey_J1J2mod-XXX,Zhu:2013_honey_J1J2mod-XXZ,Zhang:2013_honey,Gong:2013_J1J2mod-XXX,Yu:2014_honey_J1J2mod}).
The CCM technique, which we employ here, for example, yields a
critical value for $J_{2}/J_{1}$ of 0.207(3) for the QCP above which
N\'{e}el order vanishes.

Turning now to the corresponding $J_{1}$--$J_{2}$--$J_{1}^{\perp}$
honeycomb-lattice bilayer model, it is clear that at the classical
$(s \to \infty)$ level, the introduction of the NN interlayer coupling
$J_{1}^{\perp}$ is basically trivial.  Since it introduces no
additional frustration into the system, the classical N\'{e}el phase
(and the spiral phases) are completely unaffected.  All that happens
is that the NN interlayer pairs simply anti-align (for the case
$J_{1}^{\perp}>0$ considered here).  However, for the quantum versions
of the model (with the spin quantum number $s$ taking a discrete
finite value) the situation is more complicated and more interesting.
Thus, for example, when the NN interlayer coupling $J_{1}^{\perp}$ is
large enough to dominate the intralayer NN and NNN couplings, $J_{1}$
and $J_{2}$, respectively, one clearly expects the GS phase to be an
interlayer dimer VBC (IDVBC) phase, with NN spins across the two
layers forming independent spin-zero dimers.  Since the energy of each
such dimer is $-\frac{3}{4}J_{1}^{\perp}$ for our spin-$\frac{1}{2}$
model, we expect the GS energy per spin for the bilayer system in this
limit to be given by
\begin{equation}
\frac{E}{N} \xrightarrow[J_{1}^{\perp}\to \infty]{} \frac{E^{{\rm IDVBC}}}{N} = -\frac{3}{8}J_{1}^{\perp}\,.  \label{IDVBC_eq}
\end{equation}
Whereas the N\'{e}el state, and all other quasiclassical states with magnetic LRO, are gapless, since their basic excitations are Goldstone magnon modes, the IDVBC state will be gapped.  The breaking of a single spin-zero interlayer dimer to form a spin-1 NN pair is the lowest-lying excited state for the IDVBC state, and hence we expect the triplet spin gap in this limit to be given by
\begin{equation}
\Delta \xrightarrow[J_{1}^{\perp}\to \infty]{} \Delta^{{\rm IDVBC}} = J_{1}^{\perp}\,.  \label{IDVBC_triplet_spin_gap_limit-J1perp-infty_eq}
\end{equation}

Thus, in the $J_{1}$--$J_{2}$--$J_{1}^{\perp}$ model on the honeycomb
lattice, there are two different ways to destabilize the N\'{e}el LRO.
One can either increase the frustration on each layer by increasing
the strength $J_{2}$ of the NNN intralayer AFM interaction, or
increase the strength $J_{1}^{\perp}$ of the NN interlayer AFM
coupling.  While the effect of the former for the extreme quantum
case, $s=\frac{1}{2}$, has been much studied, as discussed above, the
effect of the latter has received much less attention.  It is our
intention here to apply to this model the same method, namely the CCM,
as has previously been applied to the model in the absence of
interlayer coupling $(J_{1}^{\perp} = 0)$ with considerable success.

\section{THE COUPLED CLUSTER METHOD}
\label{ccm_sec}
The CCM has been successfully applied to a wide variety
of quantum many-body systems
\cite{Fa:2004_QM-coll,Bishop:1987_ccm,Bishop:1991_TheorChimActa_QMBT,Bishop:1998_QMBT_coll},
including a large number of spin-lattice models in quantum magnetism
(and see, e.g., Refs.\
\cite{PHYLi:2012_honeycomb_J1neg,Bishop:2012_honeyJ1-J2,Bishop:2012_honey_circle-phase,Li:2012_honey_full,RFB:2013_hcomb_SDVBC,Bishop:2016_honey_grtSpins,Li:2016_honey_grtSpins,Li:2016_honeyJ1-J2_s1,DJJFarnell:2014_archimedeanLatt,Bishop:2015_honey_low-E-param,Fa:2004_QM-coll,Zeng:1998_SqLatt_TrianLatt,Bishop:2000_XXZ,Kruger:2000_JJprime,Fa:2001_SqLatt_s1,Darradi:2005_Shastry-Sutherland,Bi:2008_EPL_J1J1primeJ2_s1,Bi:2008_JPCM_J1xxzJ2xxz_s1,Bi:2009_SqTriangle,Bishop:2010_UJack,Bishop:2010_KagomeSq,Bishop:2011_UJack_GrtSpins,PHYLi:2012_SqTriangle_grtSpins,Li:2012_anisotropic_kagomeSq}.
The method affords a well-structured computational framework in which
to study a variety of candidate GS phases and their actual regimes of
stability.  The description in each case is capable of systematic
improvement via well-defined computational hierarchies of
approximations for the multispin quantum correlations incorporated.
We now describe briefly the main elements and the key features of the
CCM, and refer the interested reader to the extensive literature (and
see, e.g., Refs.\
\cite{Fa:2004_QM-coll,Bishop:1987_ccm,Bishop:1991_TheorChimActa_QMBT,Bishop:1998_QMBT_coll,Kummel:1978_ccm,Bishop:1978_ccm,Bishop:1982_ccm,Arponen:1983_ccm,Bartlett:1989_ccm,Arponen:1991_ccm,Zeng:1998_SqLatt_TrianLatt}
and references cited therein) for further details.

The CCM is a size-extensive method, and automatically provides results
in the limit of an infinitely large number of lattices spins,
$N \rightarrow \infty$, at every level of approximation, as we shall
see in more detail below.  The method is implemented in practice by
first choosing a suitable normalized reference (or model) state
$|\Phi\rangle$, with respect to which the multiparticle correlations
present in the exact GS wave function $|\Psi\rangle$ can be
incorporated systematically to higher and higher orders of
approximation as one approaches the (usually unattainable in practice)
exact limit.  As we shall see, the role of $|\Phi\rangle$ is that of a
generalized vacuum state, broadly speaking.  The exact GS ket wave
function $|\Psi\rangle$ is chosen to satisfy the intermediate
normalization condition,
$\langle\Phi|\Psi\rangle = \langle\Phi|\Phi\rangle\equiv 1$, and its
corresponding bra counterpart is defined as
$\langle\tilde{\Psi}| \equiv \langle\Psi|/\langle\Psi|\Psi\rangle$, so
that $\langle\tilde{\Psi}|\Psi\rangle = 1$.  For the present case
$|\Phi\rangle$ will be chosen to be the quasiclassical AFM N\'{e}el
state.

One of the defining features of the CCM is the specific way in which
$|\Psi\rangle$ and $\langle\tilde{\Psi}|$ are now independently
parametrized with respect to the model state $|\Phi\rangle$ in terms
of two correlation operators $S$ and $\tilde{S}$ via the exponentiated
forms,
\begin{equation}
|\Psi\rangle = e^{S}|\Phi\rangle\,; \quad \langle\tilde{\Psi}|=\langle\Phi|\tilde{S}e^{-S}\,,   \label{wave_functn_Psi_Eq}
\end{equation}
that are characteristic of the method.  Despite the fact that the
destruction correlation operator $\tilde{S}$ may be expressed formally
in terms of its creation counterpart $S$, by using Hermiticity, as
\begin{equation}
\langle\Phi|\tilde{S} = \frac{\langle\Phi|e^{S^{\dagger}}e^{S}}{\langle\Phi|e^{S^{\dagger}}e^{S}|\Phi\rangle}\,,   \label{Destruct_operator_hermiticity_Eq}
\end{equation}
the choice is made within the CCM to treat $S$ and $\tilde{S}$ as
independent operators, which will clearly satisfy Eq.\
(\ref{Destruct_operator_hermiticity_Eq}) when no approximations are made, but which
may violate this relation when truncations are made.  They are
formally decomposed as
\begin{equation}
S=\sum_{I\neq 0}{\cal{S}}_{I}C_{I}^{+}\,, \quad \tilde{S}=1 + \sum_{I \neq 0} {\cal{\tilde{S}}}_{I}C_{I}^{-}\,,      \label{sum_create_destruct_operators_Eq}
\end{equation}
where $C_{0}^{+} \equiv 1$ is defined to be the identity operator in
the many-body Hilbert space, and where the set index $I$ denotes an
appropriate complete set of single-particle configurations for all $N$
particles.  The model state $|\Phi\rangle$ is thus required to be a
cyclic (or fiducial) vector with respect to the set $\{C_{I}^{+}\}$ of
mutually commuting many-body creation operators,
\begin{equation}
[C_{I}^{+},C_{J}^{+}]=0\,, \quad \forall I,J \neq 0\,,    \label{create_destruct_operators_commute_relation_Eq}
\end{equation}
such that the set of
states $\{C_{I}^{+}|\Phi\rangle\}$ completely span the ket-state Hilbert
space.  Furthermore, we require that
\begin{equation}
\langle\Phi|C_{I}^{+} = 0 = C_{I}^{-}|\Phi\rangle\,, \quad \forall I \neq 0\,,  \label{create_destruct_operators_relations_Eq}
\end{equation}
where $C_{I}^{-} \equiv (C_{I}^{+})^{\dagger}$ is the respective
multiconfigurational destruction operator.  It is also convenient in
practice to choose the set of states $\{C_{I}^{+}|\Phi\rangle\}$ to be
orthonormal,
\begin{equation}
\langle\Phi|C_{I}^{-}C_{J}^{+}|\Phi\rangle=\delta(I,J)\,, \quad \forall I,J \neq 0\,,  \label{create_destruct_operators_orthonornmal_Eq}
\end{equation}
where $\delta(I,J)$ is a generalized Kronecker symbol.

Some important consequences immediately flow from the general CCM
parametrization of Eqs.\ (\ref{wave_functn_Psi_Eq}),
(\ref{sum_create_destruct_operators_Eq})--(\ref{create_destruct_operators_relations_Eq}).
Firstly, while explicit Hermiticity is {\it not} maintained within the
CCM, a balancing feature is that the Goldstone linked cluster theorem
{\it is} preserved, both in the exact formalism and at all levels of
approximation when truncations are made to some subset of the
set-indices $I$ in the decompositions of the correlation operators $S$
and $\tilde{S}$ expressed in Eq.\
(\ref{sum_create_destruct_operators_Eq}), the proof of which we
outline below.  As a consequence, size-extensivity is preserved in all
orders of approximation, and all thermodynamically extensive variables
such as the GS energy are guaranteed to scale linearly with particle
number $N$.  Thus, in the CCM we work from the outset in the
thermodynamic (or bulk) limit, $N \rightarrow \infty$.  Hence, there
is never any need for any finite-size scaling of our results, as is
required in many alternative techniques.  Secondly, the exponentiated
forms of the CCM parametrizations of Eq.\ (\ref{wave_functn_Psi_Eq})
also similarly guarantee the exact preservation of the important
Hellmann-Feynman theorem at all levels of approximate implementation
of the method.  This feature guarantees self-consistency among the
results for the physical parameters calculated at the same given level
of approximation.

Clearly, from Eqs.\
(\ref{wave_functn_Psi_Eq}) and (\ref{sum_create_destruct_operators_Eq}),
a knowledge of the CCM $c$-number coefficients
$\{{\cal S}_{I},{\cal \tilde{S}}_{I}\}$ suffices to determine the GS
expectation value of any physical operator.  In turn, these
coefficients are formally determined from minimizing the GS energy
expectation functional,
\begin{equation}
\bar{H} = \bar{H}({\cal S}_{I},{\cal \tilde{S}}_{I}) \equiv \langle\Phi|\tilde{S}e^{-S}He^{S}|\Phi\rangle\,,   \label{bar-H-Eq}
\end{equation}
from Eq.\ (\ref{wave_functn_Psi_Eq}), with respect to each of the
parameters $\{{\cal S}_{I},\tilde{{\cal S}}_{I};\forall I \neq 0\}$.
Extremization of $\bar{H}$ with respect to the coefficient
$\tilde{{\cal S}}_{I}$ from Eq.\
(\ref{sum_create_destruct_operators_Eq}) trivially yields the
necessary condition
\begin{equation}
\langle\Phi|C_{I}^{-}e^{-S}He^{S}|\Phi\rangle = 0\,, \quad \forall I \neq 0\,.  \label{non_linear_ket_Eq}
\end{equation}
Equation (\ref{non_linear_ket_Eq}) is manifestly a coupled set of
nonlinear equations for the set of GS ket-state coefficients
$\{{\cal S}_{I}\}$.  Formally, there are as many equations in the set
as there are unknown parameters.  Similarly, extremization of $\bar{H}$
from Eq.\ (\ref{bar-H-Eq}) with respect to the coefficient
${\cal S}_{I}$ from Eq.\ (\ref{sum_create_destruct_operators_Eq}),
leads to the corresponding set of coupled linear equations for the GS
bra-state coefficients $\{\tilde{{\cal S}}_{I}\}$,
\begin{equation}
\langle\Phi|\tilde{S}e^{-S}[H,C_{I}^{+}]e^{S}|\Phi\rangle = 0\,, \quad \forall I \neq 0\,.  \label{linear_bra_Eq}
\end{equation}
Again, once the coefficients $\{{\cal S}_{I}\}$ have been obtained
from solving Eq.\ (\ref{non_linear_ket_Eq}), they can be used as input
to Eq.\ (\ref{linear_bra_Eq}), which again formally comprises the same
number of (now linear) equations as unknown parameters
$\{\tilde{{\cal S}}_{I}\}$.

Once Eq.\ (\ref{non_linear_ket_Eq}) is satisfied, the GS energy $E$ is
simply given as the value of $\bar{H}$ from Eq.\ (\ref{bar-H-Eq}) at
the minimum, namely
\begin{equation}
E=\langle\Phi|e^{-S}He^{S}|\Phi\rangle = \langle\Phi|He^{S}|\Phi\rangle\,,  \label{E_GS_Eq}
\end{equation}
where, in the latter equation, we have also used Eqs.\
(\ref{sum_create_destruct_operators_Eq}) and
(\ref{create_destruct_operators_relations_Eq}).  Equation
(\ref{linear_bra_Eq}) may also be written in the equivalent form
\begin{equation}
\langle\Phi|\tilde{S}(e^{-S}He^{S}-E)C_{I}^{+}|\Phi\rangle = 0\,, \quad \forall I \neq 0\,,  \label{linear_bra_Eq_equivalentForm}
\end{equation}
where we have made use of Eq.\ (\ref{E_GS_Eq}).  Equation
(\ref{linear_bra_Eq_equivalentForm}) then takes the explicit form of a
set of generalized linear eigenvalue equations for the coefficients
$\{\tilde{{\cal S}}_{I}\}$, once the coefficients
$\{{\cal S}_{I}\}$ are assumed known, having been first
obtained by solving Eq.\ (\ref{non_linear_ket_Eq}).

Excited-state (ES) ket wave functions $|\Psi_{e}\rangle$ are now
parametrized in the CCM in terms of an excitation operator
\begin{equation}
X^{e} = \sum_{I \neq 0}{\cal X}_{I}^{e}C_{I}^{+}\,,   \label{excite_operator_Eq}
\end{equation}
employed linearly as
\begin{equation}
|\Psi_{e}\rangle = X^{e}e^{S}|\Phi\rangle\,.
\end{equation}
By suitably combining the GS ket-state Schr\"{o}dinger equation,
\begin{equation}
H|\Psi\rangle = E|\Psi\rangle\,; \quad \langle\tilde{\Psi}|H = E\langle\tilde{\Psi}|\,,
\end{equation}
and its ES ket-state counterpart,
\begin{equation}
H|\Psi_{e}\rangle = E_{e}|\Psi_{e}\rangle\,,
\end{equation}
we readily find the result
\begin{equation}
e^{-S}[H,X^{e}]e^{S}|\Phi\rangle = \Delta_{e}X^{e}|\Phi\rangle\,,   \label{excited_state_Eq}
\end{equation}
where we have used the fact that the operators $X^{e}$ and $S$
commute, due to their explicit defining forms in Eqs.\
(\ref{sum_create_destruct_operators_Eq}) and
(\ref{excite_operator_Eq}), together with Eq.\
(\ref{create_destruct_operators_commute_relation_Eq}), and where $\Delta_{e}$ is
defined to be the excitation energy,
\begin{equation}
\Delta_{e} = E_{e}-E\,.
\end{equation}
The ES ket-state coefficients $\{{\cal X}_{I}^{e}\}$ and the excitation
  energy $\Delta_{e}$ may then be found from solving the corresponding set of equations
\begin{equation}
\langle\Phi|C_{I}^{-}[e^{-S}He^{S},X^{e}]|\Phi\rangle = \Delta_{e}{\cal X}_{I}^{e}\,, \quad \forall I \neq 0\,,  \label{excited_state_coeff_Eq}
\end{equation}
which is readily obtained by taking the overlap of Eq.\
(\ref{excited_state_Eq}) with the state $\langle\Phi|C_{I}^{-}$, and
using both Eqs.\ (\ref{create_destruct_operators_commute_relation_Eq}) and
(\ref{create_destruct_operators_orthonornmal_Eq}) together with the
previous observation that the operators $X^{e}$ and $S$ commute.
Equation (\ref{excited_state_coeff_Eq}) again takes the form of a set
of generalized linear eigenvalue equations.

It is interesting to note at this point that the exponential terms
$e^{\pm S}$ that are a hallmark of the CCM GS and ES wave function
parametrizations, always actually enter the equations that need to be
solved for the CCM coefficients
$\{{\cal S}_{I},{\tilde{\cal S}}_{I}\}$ and $\{{\cal X}_{I}^{e}\}$
only in the specific form of the similarity transform $e^{-S}He^{S}$
of the system Hamiltonian, as may be seen explicitly from Eqs.\
(\ref{non_linear_ket_Eq}), (\ref{linear_bra_Eq_equivalentForm}), and
(\ref{excited_state_coeff_Eq}).  This may be expanded as the nested
commutator sum,
\begin{equation}
e^{-S}He^{S}=\sum_{n=0}^{\infty}\frac{1}{n!}[H,S]_{n}\,,  \label{H_similarity_transform_expansion_Eq}
\end{equation}
where the $n$-fold nested commutator $[H,S]_{n}$ is defined
iteratively for all integers $n \geq 0$ as
\begin{equation}
[H,S]_{n} \equiv [[H,S]_{n-1},S]\,; \quad [H,S]_{0}=H\,.
\end{equation}
A further key feature of the CCM parametrization of Eqs.\
(\ref{wave_functn_Psi_Eq}) and
(\ref{sum_create_destruct_operators_Eq}) is that the infinite sum in
Eq.\ (\ref{H_similarity_transform_expansion_Eq}) will generally, in
any practical implementation of the method, actually terminate (as
here) at some low finite order.  The reasons for this are due to the
facts that the Hamiltonian $H$ usually (as in the present case)
contains only finite-order multinomial terms in the corresponding set
of single-particle operators, and that all of the elements of $S$ in
its decomposition of Eq.\ (\ref{sum_create_destruct_operators_Eq})
mutually commute, as in Eq.\
(\ref{create_destruct_operators_commute_relation_Eq}).  Thus, for
example, if $H$ involves no more than $m$-body interaction terms, its
second-quantized form will contain terms with no more than $2m$
single-particle creation and annihilation operators.  As a consequence
the sum in Eq.\ (\ref{H_similarity_transform_expansion_Eq}) then
terminates exactly at the term $n=2m$, with all higher-order terms
vanishing identically.

Similarly, in the present case we note that our Hamiltonian expressed
by Eq.\ (\ref{H_eq}) is bilinear in the SU(2) spin operators
$(s_{k}^{x},s_{k}^{y},s_{k}^{z})$.  We shall describe below in detail
how in this case the operators $\{C_{I}^{+}\}$ are constructed as
products of single spin-raising operators,
$s_{k}^{+}\equiv s_{k}^{x} + is_{k}^{y}$, on various lattice sites
$k$.  The SU(2) commutation relations then immediately imply that the
sum in Eq.\ (\ref{H_similarity_transform_expansion_Eq}) terminates at
the term with $n=2$ for this case, with all nested commutators with
$n \geq 2$ vanishing identically.

Another consequence of the required mutual commutativity relation of
Eq.\ (\ref{create_destruct_operators_commute_relation_Eq}) between any pair of
operators from the set $\{C_{I}^{+}\}$ is that all non-vanishing terms
in the expansion of Eq.\ (\ref{H_similarity_transform_expansion_Eq})
for the similarity transform $e^{-S}He^{S}$ of the Hamiltonian must be
linked to the Hamiltonians.  Unlinked terms simply cannot arise due to
Eq.\ (\ref{create_destruct_operators_commute_relation_Eq}), even when any
truncation is made for the operator $S$ by curtailing the set-indices
$I$ retained in the expansion of Eq.\
(\ref{sum_create_destruct_operators_Eq}) to some selected subset.
Hence, the CCM is bound by construction to preserve the Goldstone
linked-cluster theorem (and its corollary of size extensivity) at any
level of practical implementation, as already alluded to above.

From our previous discussion it is thus clear that the {\it sole}
approximation that is needed to implement the CCM in practice is to
restrict the set of multiconfigurational set-indices $\{I\}$ that we
retain in the expansions of Eqs.\
(\ref{sum_create_destruct_operators_Eq}) and
(\ref{excite_operator_Eq}) for the GS coefficients
$\{{\cal S}_{I},\tilde{{\cal S}}_{I}\}$ and ES coefficients
$\{{\cal X}_{I}^{e}\}$, respectively, to some manageable (finite or
infinite) subset.  Any such choice of approximation will depend on the
particular model being studied and, especially, on the chosen model
state $|\Phi\rangle$ and the associated set of creation operators
$\{C_{I}^{+}\}$.  We thus describe now how such choices can be made
both for general spin-lattice models and, more particularly, for the
present bilayer model.

For an arbitrary quantum spin-lattice system a particularly simple
choice of model state $|\Phi\rangle$ is a quasiclassical
independent-spin product state.  All such states are characterized by
an independent specification of the projection of the spin on
every lattice site, along some given quantization axis (which may
itself vary from site to site).  The collinear AFM N\'{e}el state for
the present bilayer model, in which all of the spins on sites denoted
by filled circles in Fig.\ \ref{model_pattern}(a) point in a given
(arbitrary) direction, and all those on sites denoted by open circles
point in the opposing direction, is an example of such an
independent-spin product state.  It will be precisely our choice of
model state for the GS results presented in Sec.\ \ref{results_sec}.
More generally, {\it any} quasiclassical state with perfect magnetic
LRO offers a similar choice for a CCM GS model state.  It is
convenient to treat all such states in the same way.  One simple way
to do so is to make a passive rotation of each spin independently
(i.e., by choosing suitable local spin quantization axes on each
lattice site independently), such that every spin points in the same
direction, say downwards (i.e., along the negative $z_{s}$ direction).
All lattice sites thus become equivalent to one another, and all such
independent-spin product states take the universal form
$|\Phi\rangle =
|\downarrow\downarrow\downarrow\cdots\downarrow\rangle$
in their own local sets of spin quantization axes.  At the same time
such rotations are clearly just unitary spin transformations that
leave the basic SU(2) commutation relations unaltered.  The big
advantage, however, is that once such local frames are selected, all
that needs to be done to distinguish one case from another is to
rewrite the model Hamiltonian $H$ in terms of the specific choice
made.

With such a choice of local spin axes, it is evident that
$|\Phi\rangle$ is a fiducial vector with respect to a set of mutually
commuting creation operators $\{C_{I}^{+}\}$ that are chosen to be
products of single-spin raising operators $s_{k}^{+}$.  More
specifically, we choose
$C_{I}^{+} \rightarrow s_{k_{1}}^{+}s_{k_{2}}^{+} \cdots
s_{k_{n}}^{+};\,n=1,2,\cdots,2sN$,
where $s$ is the spin quantum number ($=\frac{1}{2}$, for the present
model) of all $N$ spins.  The multiconfigurational set-index $I$
correspondingly becomes a set of lattice-site indices,
$I \rightarrow \{k_{1},k_{2},\cdots,k_{n};\,n=1,2,\cdots,2sN\}$,
wherein any given site index $k_{i}$ may appear up to $2s$ times.

We shall employ here a rather general and systematic approximation
scheme for the choice of which configurations $\{I\}$ to retain in the
sums in Eq.\ (\ref{sum_create_destruct_operators_Eq}) for the
decompositions of the CCM GS correlation operators
$\{{\cal S},\tilde{\cal S}\}$, which has proven to be very powerful in
many previous applications to a diverse array of spin-lattice models.
This is the so-called localized (lattice-animal-based subsystem)
LSUB$n$ scheme.  At the $n$th level of approximation it retains all
such multispin configurations that describe clusters of spins spanning
a range of no more than $n$ contiguous lattice sites.  Contiguity of a
set of lattice sites is defined so that every site in the set is NN to
at least one other in the set (in some specified geometry).
Equivalently, in the LSUB$n$ scheme, the configurations retained are
those defined on all possible lattice animals (or polyominos) up to
size $n$.  Clearly as the truncation parameter grows without bound
($n \rightarrow \infty$) the corresponding LSUB$\infty$ approximation
becomes exact.

The effective size of the index set $\{I\}$ retained at a given
LSUB$n$ level is reduced by making use of the space- and point-group
symmetries of the lattice and the particular model state
$|\Phi\rangle$ being used, as well as any pertinent conservation laws.
For example, for the present model of Eq.\ (\ref{H_eq}) and the
N\'{e}el model state, the total $z$-component of spin,
$s_{T}^{z} \equiv \sum_{k=1}^{N}s_{k}^{z}$, is conserved (i.e.,
$s_{T}^{z} = 0$ for the N\'{e}el state), where global spin axes are
assumed.  Even so, the number $N_{f}=N_{f}(n)$ of distinct (and
nonzero) fundamental multispin-flip configurations that are retained
at a given $n$th level of LSUB$n$ approximation grows rapidly
(typically, super-exponentially) with the truncation index $n$.  For
example, for the spin-$\frac{1}{2}$ honeycomb monolayer, we have
$N_{f}(10)=6\,237$ and $N_{f}(12)=103\,097$ for the N\'{e}el GS.  By
contrast, for the present spin-$\frac{1}{2}$ honeycomb bilayer,
$N_{f}(10)=70\,118$.

For the ES calculation of the triplet spin gap, $\Delta$, the set of
LSUB$n$ multispin-flip cluster configurations $\{I\}$ retained in the
decomposition of Eq.\ (\ref{excite_operator_Eq}) for excitation
operator $X^{e}$ is, of course, different to that retained in the
corresponding decompositions of Eq.\
(\ref{sum_create_destruct_operators_Eq}) for the GS correlation
operators $(S,\tilde{S})$ at the same $n$th level of approximation.
Thus, for the calculation of the triplet spin gap based on the
N\'{e}el state as CCM model state we now retain only those
configurations $I$ that have $s_{T}^{z}=1$, compared to those that
have $s_{T}^{z}=0$ for the AFM GS calculation.  Nevertheless, both
sets of GS and ES calculations are preformed within the same LSUB$n$
scheme for consistency and to assure comparable levels of accuracy in
both.  We note that at a given LSUB$n$ level of approximation, based
on the same N\'{e}el model state, the number $N_{f}(n)$ of fundamental
CCM configurations is higher for the ES calculation than for the GS
calculation.  For example, for the spin-$\frac{1}{2}$ honeycomb
monolayer we have $N_{f}(10)=10\,497$ and $N_{f}(12)=182\,714$ for the
spin triplet ES, whereas for the corresponding bilayer case we have
$N_{f}(10)=121\,103$.

The derivation and solution of such large sets of equations clearly
require the use of both massive parallelization and supercomputing
resources.  Their derivation \cite{Zeng:1998_SqLatt_TrianLatt} also requires the
use of purpose-built, customized computer algebra packages
\cite{ccm_code}.  Whereas for the spin-$\frac{1}{2}$ honeycomb-lattice
monolayer we were able to perform LSUB$n$ calculations for the
spin-$\frac{1}{2}$ $J_{1}$--$J_{2}$ model with $n \leq 12$
\cite{Bishop:2012_honeyJ1-J2,RFB:2013_hcomb_SDVBC}, for the
corresponding present $J_{1}$--$J_{2}$--$J_{1}^{\perp}$ bilayer model
we are only able to perform LSUB$n$ calculations with $n \leq 10$, due
to the substantially increased numbers $N_{f}(n)$ of fundamental
configurations in this case, as illustrated above.

Whereas the GS energy $E$ can, uniquely, be calculated from a
knowledge of the CCM creation coefficients $\{{\cal S}_{I}\}$ alone,
as from Eq.\ (\ref{E_GS_Eq}), any other GS quantity requires also a
knowledge of the corresponding destruction coefficients
$\{\tilde{{\cal S}}_{I}\}$.  For example, we also calculate here the N\'{e}el
magnetic order parameter $M$.  This is defined to be the average
on-site GS magnetization using the N\'{e}el state as the CCM model
state $|\Phi\rangle$,
\begin{equation}
M = -\frac{1}{N}\sum_{k=1}^{N}\langle\Phi|\tilde{S}e^{-S}s_{k}^{z}e^{S}|\Phi\rangle\,,  \label{M_Eq}
\end{equation}
in terms of the local rotated spin-coordinate frames described previously.

The last step in our CCM calculational procedure is to extrapolate the
corresponding sequence of LSUB$n$ results for any GS or ES quantity
that we have computed to the LSUB$\infty$ limit
$(n \rightarrow \infty)$ in which all relevant multispin-flip
configurations are retained, and the method hence becomes exact.  From
our previous description of the method it should be clear that this last step is the {\it sole} approximation made in the entire procedure.  Despite the fact that, so far as we know, there exist no exact results for performing such extrapolations, a great deal of practical experience has by now been accumulated from the many applications of the CCM that have already been made to a wide variety of quantum spin-lattice systems.  For example, for the GS energy per spin, $E/N$, there exists the highly accurate and very well tested extrapolation scheme (and see, e.g., Refs.\ \cite{DJJF:2011_honeycomb,PHYLi:2012_honeycomb_J1neg,Bishop:2012_honeyJ1-J2,Bishop:2012_honey_circle-phase,Li:2012_honey_full,RFB:2013_hcomb_SDVBC,Bishop:2016_honey_grtSpins,Li:2016_honey_grtSpins,Li:2016_honeyJ1-J2_s1,DJJFarnell:2014_archimedeanLatt,Bishop:2015_honey_low-E-param,Fa:2004_QM-coll,Bishop:2000_XXZ,Kruger:2000_JJprime,Fa:2001_SqLatt_s1,Darradi:2005_Shastry-Sutherland,Bi:2008_EPL_J1J1primeJ2_s1,Bi:2008_JPCM_J1xxzJ2xxz_s1,Bi:2009_SqTriangle,Bishop:2010_UJack,Bishop:2010_KagomeSq,Bishop:2011_UJack_GrtSpins,PHYLi:2012_SqTriangle_grtSpins,Li:2012_anisotropic_kagomeSq})
\begin{equation}
\frac{E(n)}{N} = e_{0}+e_{1}n^{-2}+e_{2}n^{-4}\,,     \label{extrapo_E}
\end{equation}
from fits to which we can obtain the LSUB$\infty$ value $e_{0}$.
As one would expect {\it a priori}, the GS expectation values of other
physical operators converge more slowly than does the energy [i.e.,
with leading exponents in their extrapolation schemes comparable to
Eq.\ (\ref{extrapo_E}) that take values less than 2].  In particular,
for the order parameter $M$ of Eq.\ (\ref{M_Eq}), it has been well
documented that the corresponding exponent depends sensitively on the
degree of frustration present in the model.  Not surprisingly perhaps,
the exponent is smaller for situations with the highest frustration.

More explicitly, for models with only little or zero frustration, an
extrapolation scheme for $M$ with a leading power $n^{-1}$ (i.e., with
leading exponent equal to 1),
\begin{equation}
M(n) = m_{0}+m_{1}n^{-1}+m_{2}n^{-2}\,,   \label{M_extrapo_standard}
\end{equation}
has been found to hold well and to give highly accurate results for
many systems (and see, e.g., Refs.\
\cite{PHYLi:2012_honeycomb_J1neg,Bishop:2012_honeyJ1-J2,Bishop:2012_honey_circle-phase,RFB:2013_hcomb_SDVBC,DJJFarnell:2014_archimedeanLatt,Bishop:2000_XXZ,Kruger:2000_JJprime,Fa:2001_SqLatt_s1,Darradi:2005_Shastry-Sutherland,Bi:2009_SqTriangle,Bishop:2010_UJack,Bishop:2010_KagomeSq,Bishop:2011_UJack_GrtSpins}).
Again from such a fit we obtain the extrapolated LSUB$\infty$ estimate
$m_{0}$ for $M$.  On the other hand, for systems that are close to a
QCP, or for phases whose magnetic order parameter $M$ is either zero
or very small, the scaling ansatz of Eq.\ (\ref{M_extrapo_standard})
does not fit so well.  A forced fit then tends to overestimate the
correct LSUB$\infty$ extrapolant.  As a consequence it also tends to
predict a value for the critical strength of the frustrating
interaction, which is responsible for driving the respective phase
transition, that is too large.  In such cases a more accurate scaling
ansatz is always found to be (and see, e.g., Refs.\ \cite{DJJF:2011_honeycomb,PHYLi:2012_honeycomb_J1neg,Bishop:2012_honeyJ1-J2,Bishop:2012_honey_circle-phase,Li:2012_honey_full,RFB:2013_hcomb_SDVBC,Li:2016_honey_grtSpins,Li:2016_honeyJ1-J2_s1,DJJFarnell:2014_archimedeanLatt,Bi:2008_EPL_J1J1primeJ2_s1,Bi:2008_JPCM_J1xxzJ2xxz_s1,Li:2012_anisotropic_kagomeSq})
\begin{equation}
M(n) = \mu_{0}+\mu_{1}n^{-1/2}+\mu_{2}n^{-3/2}\,,   \label{M_extrapo_frustrated}
\end{equation}
from which $\mu_{0}$ gives the respective extrapolated LSUB$\infty$ value for $M$.

\begin{figure*}[t]
\begin{center}
\mbox{
\hspace{-1.0cm}
\subfigure[]{\scalebox{0.3}{\includegraphics[angle=270]{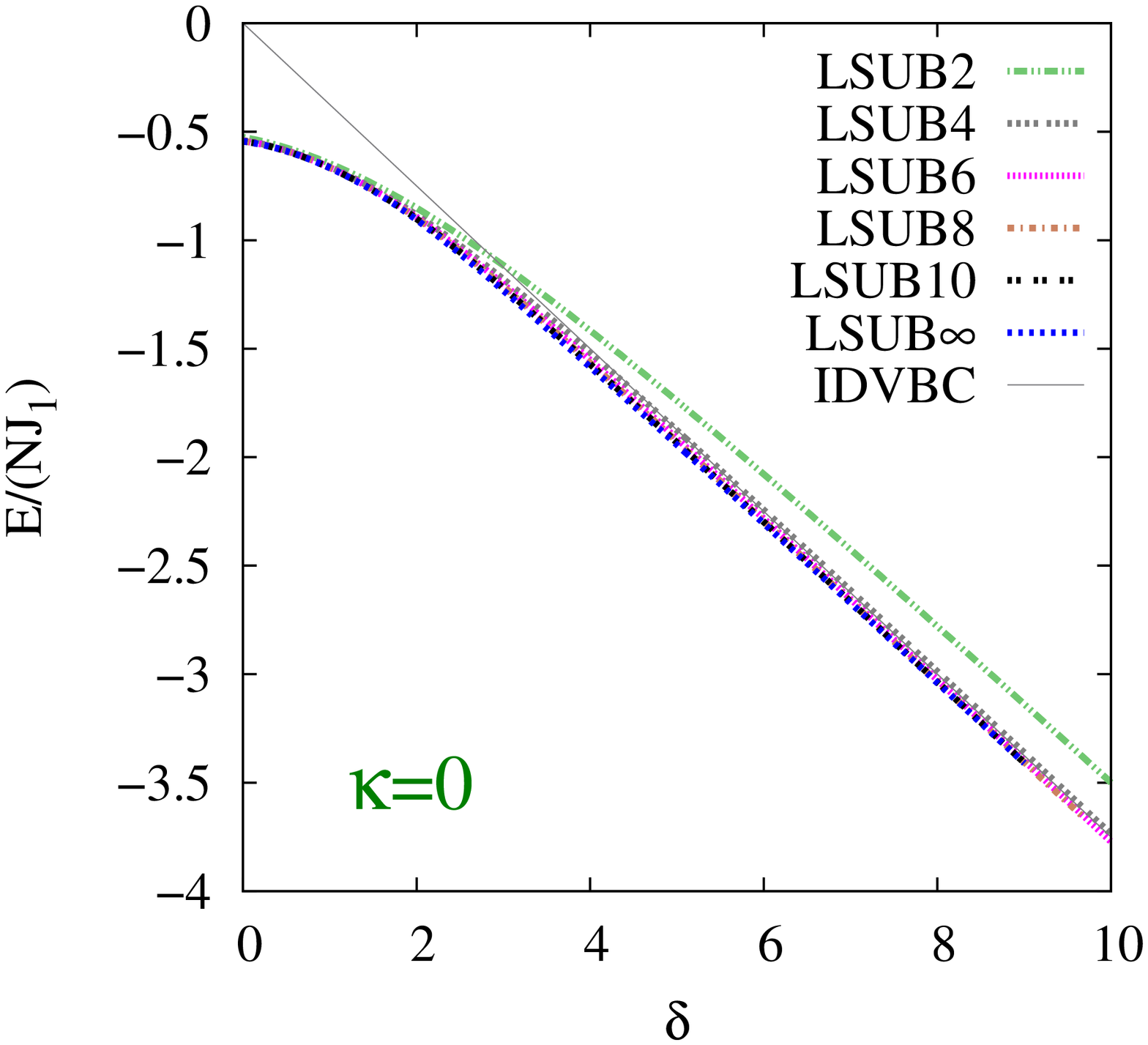}}}
\hspace{-1.8cm}
\subfigure[]{\scalebox{0.3}{\includegraphics[angle=270]{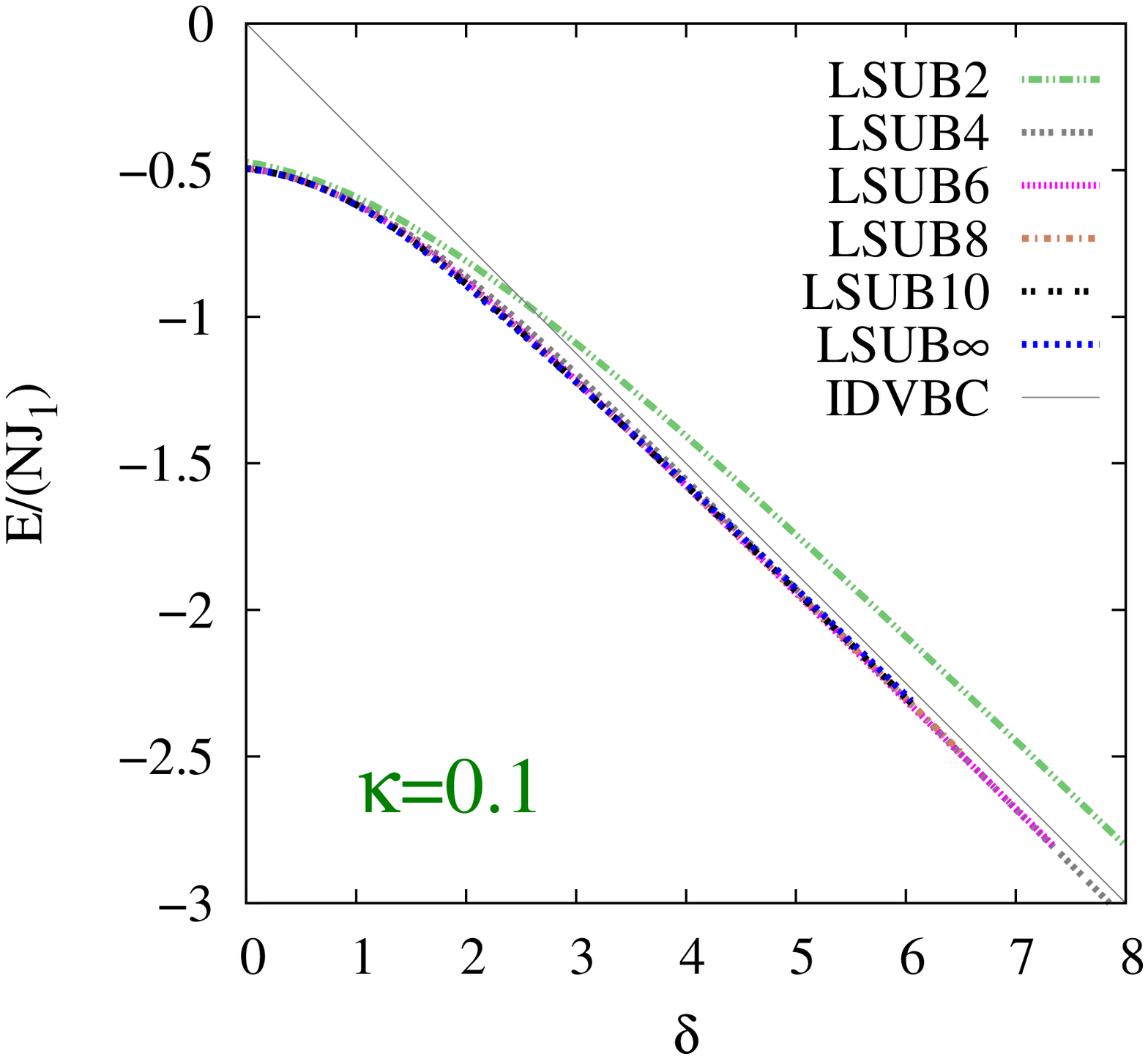}}}
\hspace{-1.8cm}
\subfigure[]{\scalebox{0.3}{\includegraphics[angle=270]{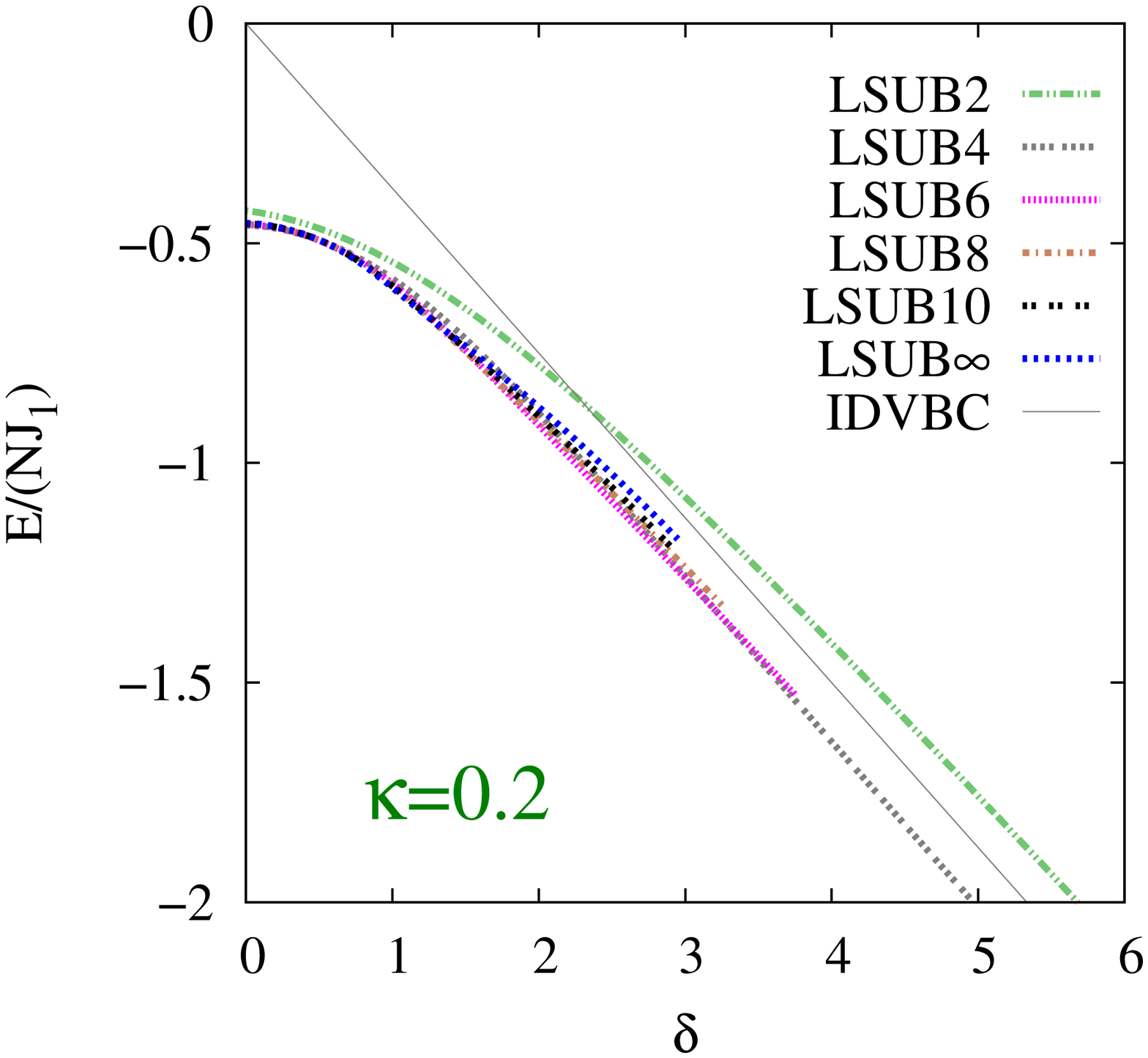}}}
}
\caption{CCM results for the GS energy per spin $E/N$ (in units of $J_{1}$) versus the scaled interlayer exchange coupling constant, $\delta \equiv J_{1}^{\perp}/J_{1}$, for the spin-$\frac{1}{2}$
    $J_{1}$--$J_{2}$--$J_{1}^{\perp}$ model on the bilayer honeycomb lattice (with
    $J_{1}>0$), for three selected values of the intralayer frustration parameter, $\kappa \equiv J_{2}/J_{1}$:  (a) $\kappa=0$, (b) $\kappa=0.1$, and (c) $\kappa=0.2$.  Results based on the N\'{e}el state as CCM model state are shown in LSUB$n$ approximations with $n=2,4,6,8,10$, together with the corresponding LSUB$\infty$ extrapolated results using Eq.\ (\ref{extrapo_E}) and the data sets $n=\{4,6,8,10\}$.   For comparison, we also show the corresponding interlayer dimer VBC result of Eq.\ (\ref{IDVBC_eq}).}
\label{E_raw_extrapo_fix-J2}
\end{center}
\end{figure*}  

Finally, a CCM extrapolation scheme with a leading power of $n^{-1}$
has also been found to give an excellent fit to the LSUB$n$
results $\Delta(n)$ for the spin gap (and see, e.g., Refs.\
\cite{Li:2016_honeyJ1-J2_s1,Bishop:2015_honey_low-E-param,Kruger:2000_JJprime,Richter:2015_ccm_J1J2sq_spinGap,Bishop:2015_J1J2-triang_spinGap}),
\begin{equation}
\Delta(n) = d_{0}+d_{1}n^{-1}+d_{2}n^{-2}\,,   \label{Eq_spin_gap}
\end{equation}
from which the corresponding LSUB$\infty$ estimate $d_{0}$ for the
spin gap $\Delta$ has been successfully extracted for a wide variety
of spin-lattice models.  

Clearly, to obtain robust fits, it is preferable to use at least four
LSUB$n$ data points in using the extrapolation schemes of Eqs.\
(\ref{extrapo_E})--(\ref{Eq_spin_gap}), since each contains three
fitting parameters.  Occasionally this may be inappropriate for
reasons we describe.  In such cases it may also be suitable to use a
wholly unbiased extrapolation scheme in which the leading exponent is
itself also a fitting parameter.  Thus, the LSUB$n$ approximants
$P(n)$ for a physical parameter $P$ are then extrapolated to give the
LSUB$\infty$ estimate $p_{0}$ via the unbiased scheme,
\begin{equation}
P(n) = p_{0}+p_{1}n^{-\nu}\,,   \label{Eq_exponFit}
\end{equation}
in which the three parameters $p_{0}$, $p_{1}$, and $\nu$ are all
treated as quantities to be fitted.

Finally, we note that we may always perform an unbiased pre-fit of the
form of Eq.\ (\ref{Eq_exponFit}) to any CCM LSUB$n$ sequence of
results for an arbitrary physical parameter, whether or not four or
more data points are available.  In this way one may first check the
(approximate) value of the exponent $\nu$, before using one of the
above schemes of Eqs.\ (\ref{extrapo_E})--(\ref{Eq_spin_gap}), for
example, which will almost always then lead to better extrapolated
results due to the addition of the next-to-leading-order correction to
the leading-order term.  In practice, such a pre-fit to the results
for the order parameter $M$ for example, essentially always leads to a
clear choice between the fits of Eqs.\ (\ref{M_extrapo_standard}) and
(\ref{M_extrapo_frustrated}).  It is in this way that the
extrapolation scheme for any physical parameter is matched in practice
to a particular regime for the Hamiltonian under study.

\section{RESULTS}
\label{results_sec}
We first show in Fig.\ \ref{E_raw_extrapo_fix-J2} our results for the
GS energy per spin as a function of the scaled interlayer exchange
coupling constant, $\delta \equiv J_{1}^{\perp}/J_{1}$, for three
different values of the interlayer frustration parameter,
$\kappa \equiv J_{2}/J_{1}$.
For each value of $\kappa$ we display the results based on the
N\'{e}el state as CCM model state at LSUB$n$ approximation levels
$n=2,4,6,8,10$.  We also show the corresponding LSUB$\infty$
extrapolated values ($e_{0}/J_{1}$) obtained from using Eq.\
(\ref{extrapo_E}) together with the data sets $n=\{4,6,8,10\}$ as
input.  We see that in each case the convergence is extremely rapid as
the truncation index $n$ is increased.  We note too that, as is always
the case in practical applications of the CCM, the LSUB$n$ results
for all finite values of the truncation index $n$ extend beyond the
actual N\'{e}el transition point out to some critical value beyond
which no real solution to the CCM equations exists.  As is also
usually true, the natural termination point for the solution tracked
for each LSUB$n$ set of GS equations, appears to converge uniformly as
$n$ is increased to the corresponding QCP at which N\'{e}el order
vanishes.  These termination points for the GS CCM equations at each
LSUB$n$ level of approximation are always direct manifestations of the
respective QCP present in the physical system being studied, as has
been well described and documented in many previous applications of
the method (and see, e.g., Refs.\
\cite{PHYLi:2012_honeycomb_J1neg,Bishop:2012_honeyJ1-J2,RFB:2013_hcomb_SDVBC,Fa:2004_QM-coll}
and references cited therein).  Furthermore, each CCM LSUB$n$ solution
extends slightly into the unphysical regime, beyond the respective
LSUB$\infty$ QCP, with the extent of the unphysical regime reducing
monotonically to zero as the LSUB$n$ truncation index $n$ is increased
to the exact $n \rightarrow \infty$ limit.

\begin{figure*}[t]
\begin{center}
\mbox{
\hspace{-1.0cm}
\subfigure[]{\scalebox{0.3}{\includegraphics[angle=270]{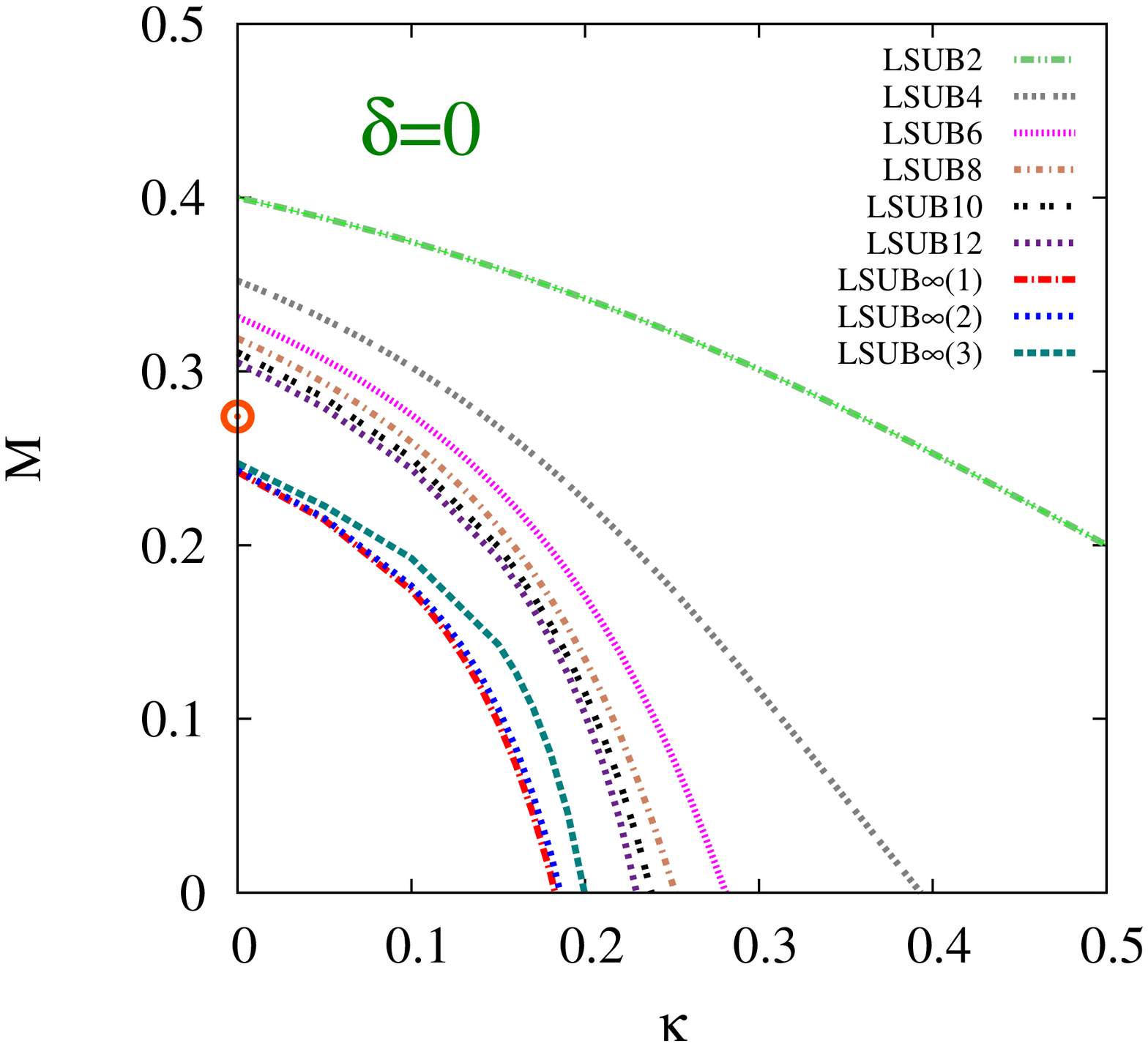}}}
\hspace{-1.8cm}
\subfigure[]{\scalebox{0.3}{\includegraphics[angle=270]{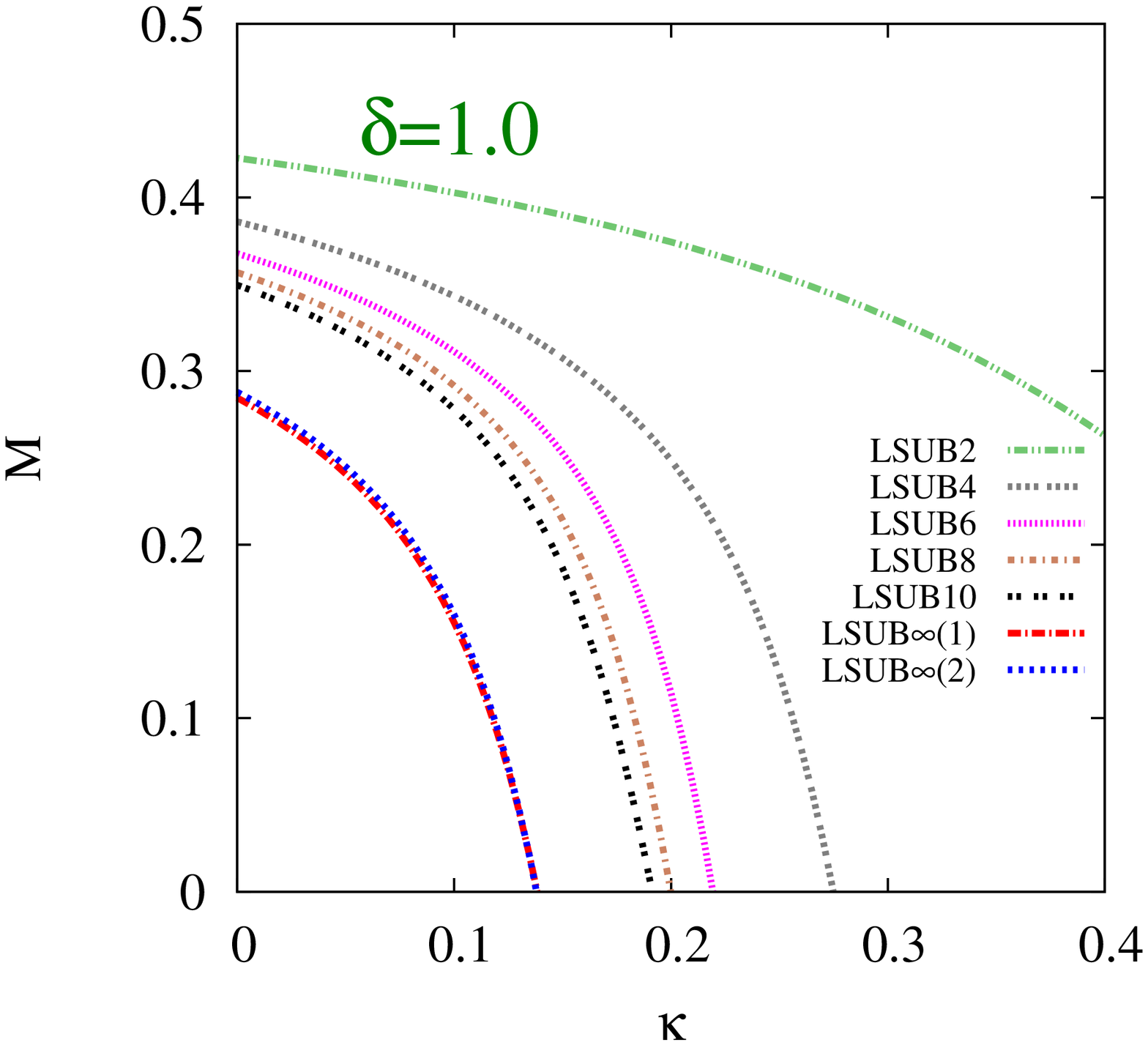}}}
\hspace{-1.8cm}
\subfigure[]{\scalebox{0.3}{\includegraphics[angle=270]{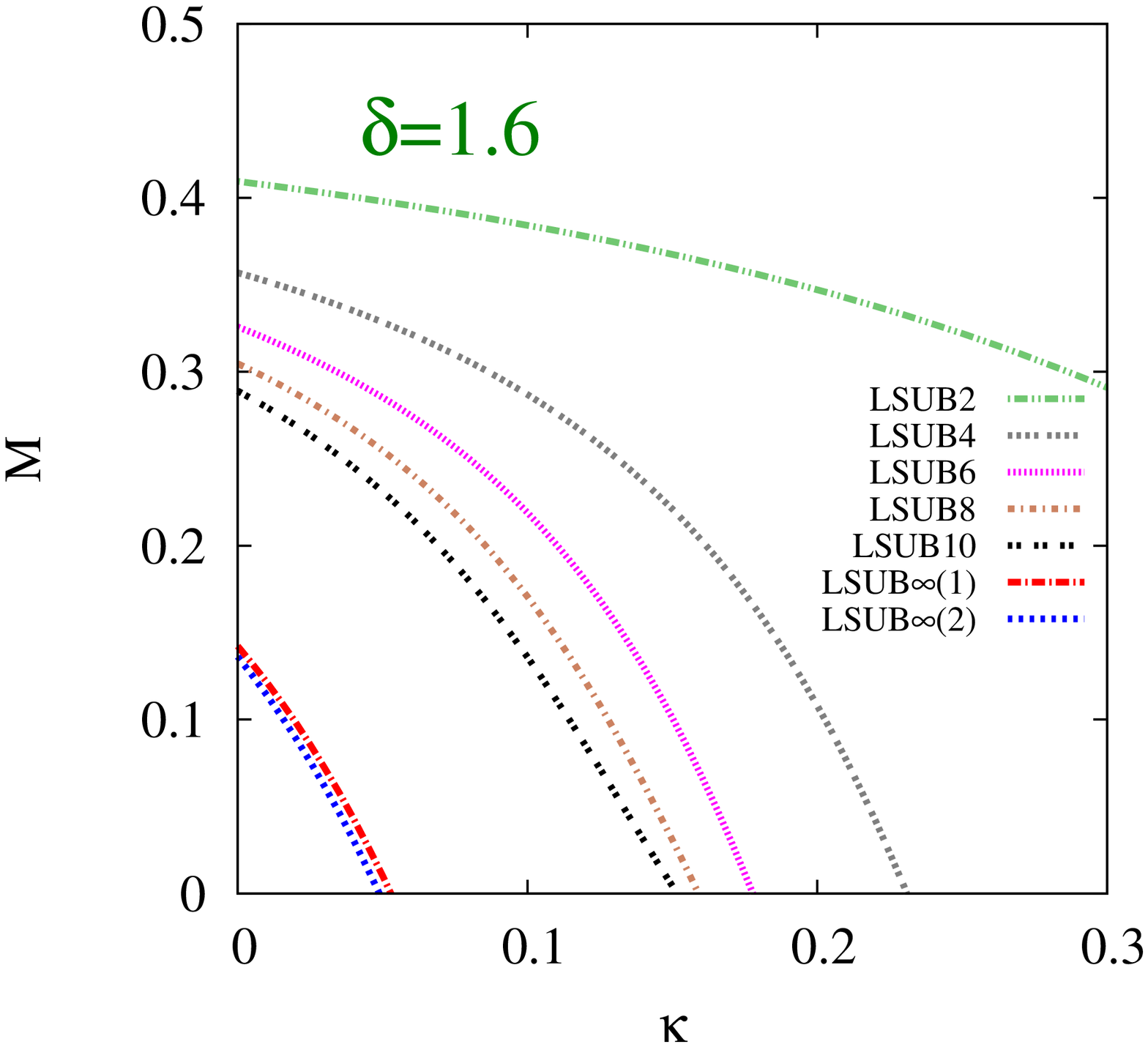}}}
}
\caption{CCM results for the GS magnetic order parameter $M$ versus
  the intralayer frustration parameter,
  $\kappa \equiv J_{2}/J_{1}$, for the spin-$\frac{1}{2}$
  $J_{1}$--$J_{2}$--$J_{1}^{\perp}$ model on the bilayer honeycomb
  lattice (with $J_{1}>0$), for three selected values of the scaled
  interlayer exchange coupling constant, $\delta \equiv J_{1}^{\perp}/J_{1}$:
  (a) $\delta=0$, (b) $\delta=1.0$, and (c) $\delta=1.6$.  Results
  based on the N\'{e}el state as CCM model state are shown in LSUB$n$
  approximations with $n=2,4,6,8,10$ (and also with $n=12$ for the
  special case of the $J_{1}$--$J_{2}$ monolayer, i.e., when
  $\delta=0$), together with various corresponding LSUB$\infty(i)$
  extrapolated results using Eq.\ (\ref{M_extrapo_frustrated}) and the
  respective data sets $n=\{2,6,10\}$ for $i=1$, $n=\{4,6,8,10\}$ for
  $i=2$, and $n=\{4,8,12\}$ for $i=3$ (for the case $\delta=0$ only).
  In Fig.\ \ref{M_raw_extrapo_fix-J1perp}(a), rather than crowd the
  figure with additional full curves based on (the largely
  inappropriate) Eq.\ (\ref{M_extrapo_standard}), we show with the
  circle ($\bigcirc$) symbol the corresponding extrapolated value
  using Eq.\ (\ref{M_extrapo_standard})
  and the data set $n=\{4,6,8,10\}$ for the single point $\delta=0=\kappa$, where this extrapolation scheme {\it is} the appropriate one.}
\label{M_raw_extrapo_fix-J1perp}
\end{center}
\end{figure*}  

Before proceeding to our results for the magnetic order parameter and
the triplet spin gap of the model, we pause to consider the accuracy
of our results.  One way to do so is the focus on the case
$\delta = 0 = \kappa$ fo the pure spin-$\frac{1}{2}$ Heisenberg
antiferromagnet (HAFM) on a honeycomb monolayer (i.e., with NN
isotropic AFM Heisenberg exchange interactions only) since, for this
unfrustrated case only, we may also compare our results with those of
large-scale quantum Monte Carlo (QMC) simulations that are available.
For this case, for example, our extrapolated LSUB$\infty$ results
$e_{0}$ based on Eq.\ (\ref{extrapo_E}) are $E/N=-0.54473(2)J_{1}$ using
the LUSB$n$ data set $n=\{4,6,8,10\}$, and $E/N=-0.54466(1)J_{1}$
using the corresponding set $n=\{6,8,10,12\}$.  In both cases the
errors cited are simply those associated with the respective fits.
The two extrapolations are in clear good agreement with one another.
They also agree extremely well with the result $E/N=-0.54455(2)J_{1}$
obtained from a large-scale, continuous Euclidean time QMC algorithm
\cite{Low:2009_honey}, and where this infinite-lattice result is based
on extrapolating the QMC estimates for finite-sized $L \times L$
lattices with $16 \leq L \leq 36$.

It has been observed previously \cite{Li:2016_honeyJ1-J2_s1} that for
the $J_{1}$--$J_{2}$ model on the honeycomb monolayer (i.e., with
$\delta=0$) there is a noticeable $(4m-2)/4m$ staggering effect in
some of the CCM LSUB$n$ sequence of results.  For example, in the
spin-1 case this is even strong enough that for the N\'{e}el magnetic
order parameter the corresponding curves with $n=4$ and $n=6$ actually
cross one another at a value of the intralayer frustration parameter
$\kappa \approx 0.2$.  Clearly, in such cases, the two separate
LSUB$n$ sequences of results (for a fixed value of $\kappa$) with
$n=4m$ and $n=4m-2$, respectively, tend to converge differently from
each other for all positive integral values of $m$, although both
sequences themselves converge monotonically.  To test for this effect
here we have also extrapolated our results for the unfrustrated
monolayer case (i.e., with $\kappa = 0 = \delta$) separately for the
two sequences, since in this case alone are we also able to perform
LSUB12 calculations.  For the GS energy per spin for this case we
obtain $E/N=-0.55473J_{1}$ using Eq.\ (\ref{extrapo_E}) with the
LSUB$n$ data set $n=\{2,6,10\}$, and $E/N=-0.55468J_{1}$ with the
corresponding set $n=\{4,8,12\}$.  The level of agreement is again
excellent.

We turn next to our results for the magnetic order parameter $M$ for
the N\'{e}el state.  Firstly, we show in Fig.\
\ref{M_raw_extrapo_fix-J1perp}(a) results for the case of the
spin-$\frac{1}{2}$ honeycomb-lattice monolayer (i.e., with
$\delta=0$), as a function of the intralayer frustration parameter
$\kappa \equiv J_{2}/J_{1}$.
In this case alone we are able to perform CCM LSUB$n$ calculations up
to values $n=12$ of the truncation parameter, and hence we can again
investigate whether the $(4m-2)/4m$ staggering effect is appreciable
in our LSUB$n$ results for $M$.  While the effect is not immediately
evident in the pattern of the raw LSUB$n$ data, some effect is
noticeable in the extrapolated results, particularly near the critical
point at which the N\'{e}el LRO vanishes (i.e., where
$M \rightarrow 0$).  

\begin{figure*}[t]
\begin{center}
\mbox{
\hspace{-1.0cm}
\subfigure[]{\scalebox{0.3}{\includegraphics[angle=270]{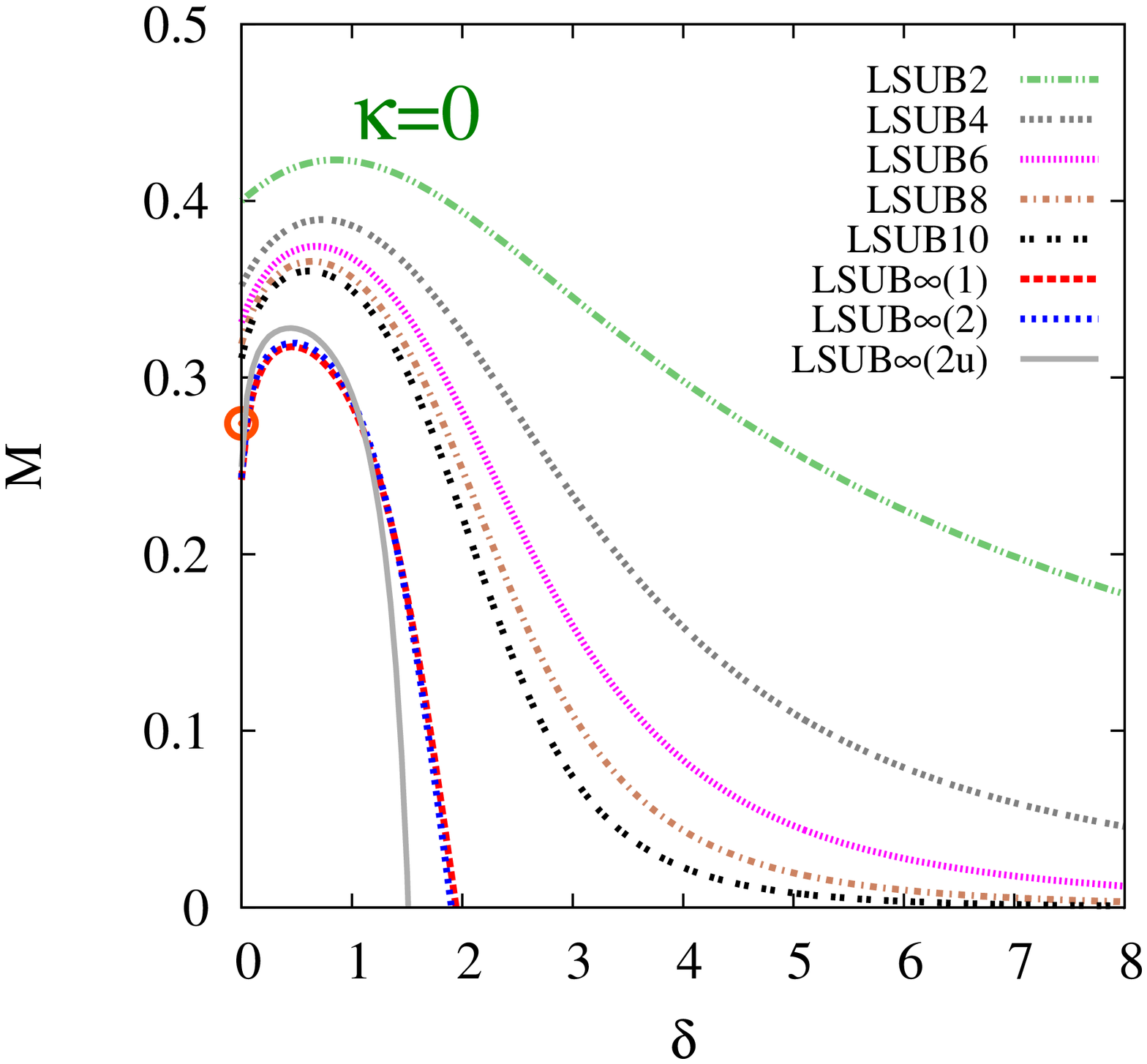}}}
\hspace{-1.8cm}
\subfigure[]{\scalebox{0.3}{\includegraphics[angle=270]{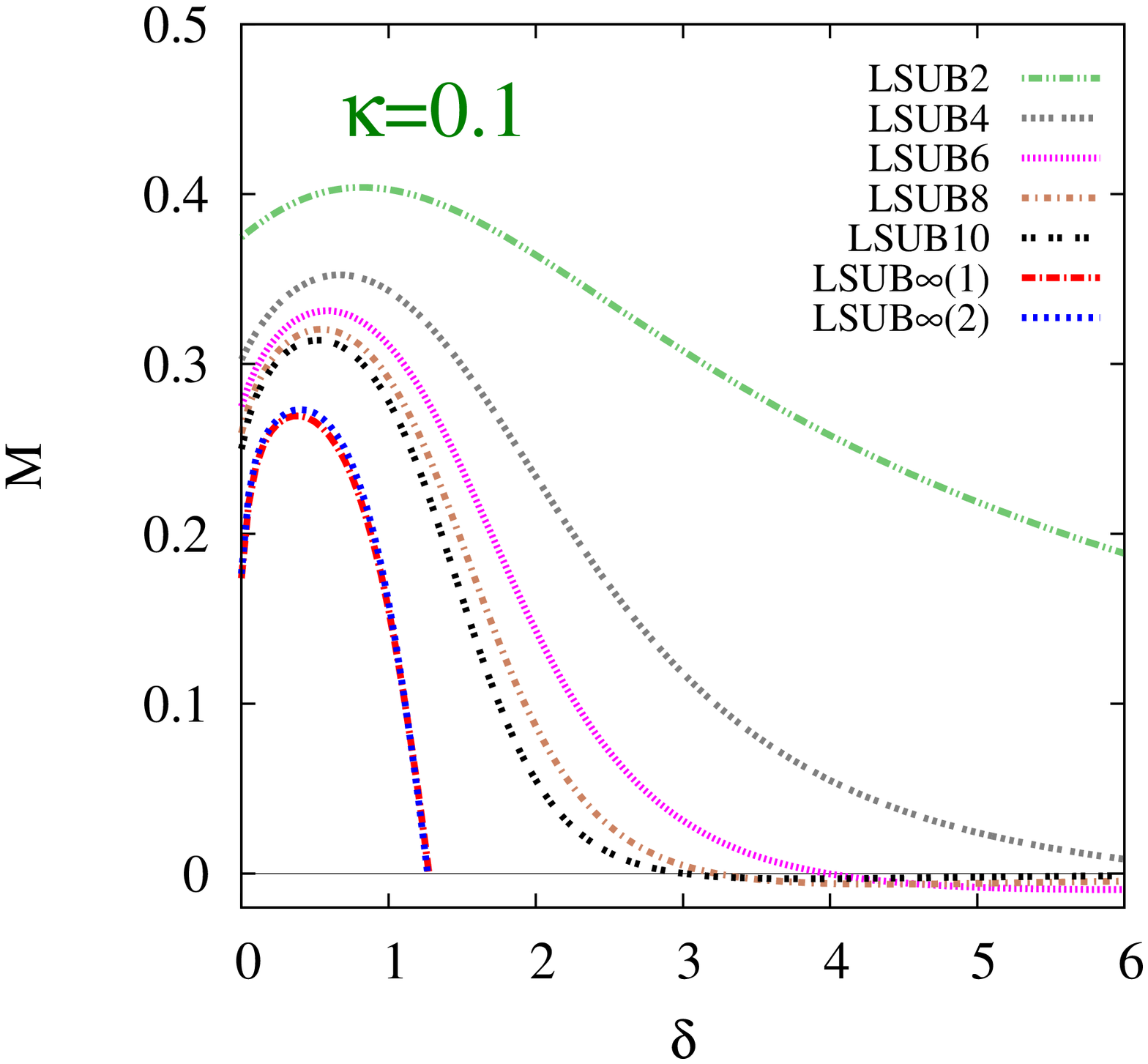}}}
\hspace{-1.8cm}
\subfigure[]{\scalebox{0.3}{\includegraphics[angle=270]{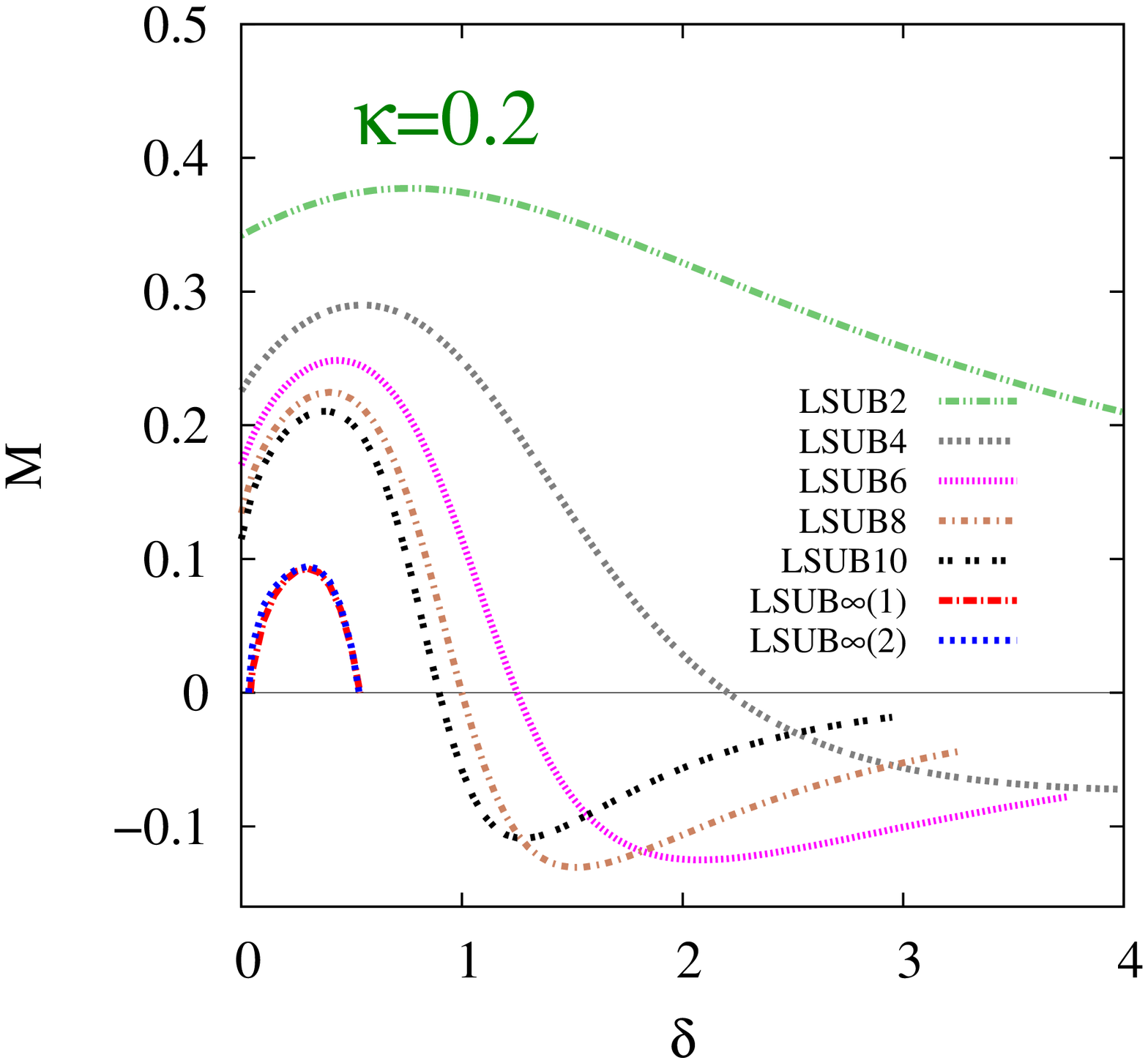}}}
}
\caption{CCM results for the GS magnetic order parameter $M$ versus
  the scaled interlayer exchange coupling constant,
  $\delta \equiv J_{1}^{\perp}/J_{1}$, for the spin-$\frac{1}{2}$
  $J_{1}$--$J_{2}$--$J_{1}^{\perp}$ model on the bilayer honeycomb
  lattice (with $J_{1}>0$), for three selected values of the
  intralayer frustration parameter, $\kappa \equiv J_{2}/J_{1}$: (a)
  $\kappa=0$, (b) $\kappa=0.1$, and (c) $\kappa=0.2$.  Results based
  on the N\'{e}el state as CCM model state are shown in LSUB$n$
  approximations with $n=2,4,6,8,10$, together with two corresponding
  LSUB$\infty(i)$ extrapolated results using Eq.\
  (\ref{M_extrapo_frustrated}) and the respective data sets
  $n=\{2,6,10\}$ for $i=1$ and $n=\{4,6,8,10\}$ for $i=2$.  In Fig.\
  \ref{M_raw_extrapo_fix-J2}(a), we also show the corresponding
  LSUB$\infty(2u)$ extrapolated result based on the unbiased scheme of
  Eq.\ (\ref{Eq_exponFit}) and the data set $n=\{4,6,8,10\}$.  Furthermore in
  Fig.\ \ref{M_raw_extrapo_fix-J2}(a), rather than crowd the figure
  with additional full curves based on (the largely inappropriate)
  Eq.\ (\ref{M_extrapo_standard}), we show with the circle
  ($\bigcirc$) symbol the corresponding extrapolated value using Eq.\
  (\ref{M_extrapo_standard}) and the data set $n=\{4,6,8,10\}$ for the
  single point $\kappa=0=\delta$, where this extrapolation scheme {\it
    is} the appropriate one.}
\label{M_raw_extrapo_fix-J2}
\end{center}
\end{figure*}

Thus, in Fig.\ \ref{M_raw_extrapo_fix-J1perp}(a) we show three
separate LSUB$\infty$ extrapolations, all based on the scheme of Eq.\
(\ref{M_extrapo_frustrated}), using the respective data sets
$n=\{2,6,10\}$, $n=\{4,6,8,10\}$, and $n=\{4,8,12\}$.  While the first
two extrapolations agree very closely with one another over the whole
range of values of the intralayer frustration parameter $\kappa$ for
which $M > 0$, the latter extrapolation does differ somewhat from the
other two, particularly near the QCP at which $M \rightarrow 0$.
While the extrapolation scheme of Eq.\ (\ref{M_extrapo_frustrated}) is
certainly appropriate for the case where the frustration is
appreciable, particularly for systems that are close to a QCP, the
scheme of Eq.\ (\ref{M_extrapo_standard}) is valid for unfrustrated
(or only slightly frustrated) systems, such as at the point
$\delta = 0 = \kappa$, and we show by the circle ($\bigcirc$) symbol
in Fig.\ \ref{M_raw_extrapo_fix-J1perp}(a) the value of $M$ at this
point using Eq.\ (\ref{M_extrapo_standard}) and the data set
$n=\{4,6,8,10\}$.  The value so obtained is $M=0.2741(1)$ where,
again, the error is simply that associated with the fit.
Corresponding values using Eq.\ (\ref{M_extrapo_standard}) and other
LSUB$n$ data set as input are $M=0.2729(5)$ using $n=\{6,8,10,12\}$,
$M=0.2761$ using $n=\{2,6,10\}$, $M=0.2733$ using $n=\{4,8,12\}$, and
$M=0.2715$ using $n=\{8,10,12\}$.  There is clearly a small sensitivity
to the data set used, which results in an overall error of about
$0.5\%$ for the value obtainable.  Once again, our CCM results may be
compared with those from two separate QMC estimates that have been
performed in this limiting unfrustrated case, namely, $M=0.2681(8)$
\cite{Low:2009_honey} and $M=0.26882(3)$ \cite{Jiang:2012_honey}.  The
agreement with our CCM results is again good.

The sensitivity of the extrapolated results for the order parameter
$M$ is, however, greater for cases where frustration is present, as
can clearly be observed from Fig.\ \ref{M_raw_extrapo_fix-J1perp}(a),
particularly in the region near the critical point where N\'{e}el LRO
melts.  Thus, for the honeycomb monolayer (i.e., when $\delta = 0$),
the corresponding LSUB$\infty$ estimates for the point $\kappa_{c}$ at
which $M \rightarrow 0$, all based on the extrapolation scheme of Eq.\
(\ref{M_extrapo_frustrated}) but with different LSUB$n$ data sets as
input are, for example, $\kappa_{c}=0.183$ using $n=\{2,6,10\}$ and
$\kappa_{c}=0.186$ based on $n=\{4,6,8,10\}$.  However, inclusion of
the LSUB12 results gives somewhat higher estimates.  For example, we
find $\kappa_{c}=0.207$ based on LSUB$n$ results with
$n=\{6,8,10,12\}$, $\kappa_{c}=0.200$ based on $n=\{4,8,12\}$, and
$\kappa_{c}=0.204$ based on $n=\{8,10,12\}$.  Clearly, in this case,
our estimates for $\kappa_{c}$ have an associated overall error of
around $5\%$.

Corresponding results to those shown for the monolayer $(\delta=0)$ in
Fig.\ \ref{M_raw_extrapo_fix-J1perp}(a) are shown in Figs.\
\ref{M_raw_extrapo_fix-J1perp}(b) and
\ref{M_raw_extrapo_fix-J1perp}(c) for the bilayer for the two cases
$\delta=1.0$ and $\delta=1.6$.  It is evident that the critical value,
$\kappa_{c}(\delta)$, at which N\'{e}el LRO melts, decreases as the
strength $\delta \equiv J_{1}^{\perp}/J_{1}$ of the interlayer coupling
  increases, at least for values of $\delta$ above some lower critical value.  We return to this point later.  The effect of the interlayer coupling is also shown
  separately in Fig. \ref{M_raw_extrapo_fix-J2} for three different
  values of the intralayer frustration parameter,
  $\kappa \equiv J_{2}/J_{1}$.
Firstly, in Fig.\ \ref{M_raw_extrapo_fix-J2}(a), we show our CCM
LSUB$n$ results with $n \leq 10$ for the N\'{e}el order parameter $M$
for the case of zero intralayer frustration (i.e., $\kappa=0$), where
only NN interactions are present.  Interestingly, as $\delta$ is first
increased from zero, the effect of interlayer AFM NN coupling is to
increase the order parameter $M$, and hence to enhance the stability
of N\'{e}el LRO.  The effect reaches a maximum at each level of
approximation at a value $\delta \approx 0.5$.  Once $\delta$ is
increased further, however, N\'{e}el order begins to reduce, and in
each LSUB$n$ approximation shown the order parameter $M$ tends to zero
asymptotically as $\delta$ continues to increase.  The relative shapes
of the LSUB$n$ curves is particularly interesting, with clear evidence
that as the truncation index $n$ is increased the asymptotic vanishing
of the order parameter becomes appreciably sharper.  

We also show in Fig.\ \ref{M_raw_extrapo_fix-J2}(a) two corresponding
LSUB$\infty$ extrapolations for $M$, both based on the scheme of Eq.\
(\ref{M_extrapo_frustrated}), but using respective LSUB$n$ data sets
with $n=\{2,6,10\}$ and $n=\{4,6,8,10\}$ as input.  Both
extrapolations agree with each other very closely, with the
corresponding estimates for the point $\delta_{c}^{>}$ at which
$M \rightarrow 0$ being $\delta_{c}^{>}=1.944$ based on the set
$n=\{2,6,10\}$, and $\delta_{c}^{>}=1.902$ based on the set
$n=\{4,6,8,10\}$.  We also show for comparison purposes in Fig.\
\ref{M_raw_extrapo_fix-J2}(a) the corresponding extrapolated result
based on the unbiased scheme of Eq.\ (\ref{Eq_exponFit}), and using
the LSUB$n$ data set $n=\{4,6,8,10\}$ as input, denoted as the
LSUB$\infty(2u)$ curve.  The overall level of agreement between the
corresponding LSUB$\infty(2)$ and LSUB$\infty(2u)$ extrapolations is
reasonable.  While it is excellent for smaller values of the scaled
interlayer coupling constant, $0 < \delta \lesssim 1$, there is a
greater degree of divergence at larger values, $\delta \gtrsim 1$.
For example, the LSUB$\infty(2u)$ extrapolation yields a value
$\delta_{c}^{>}=1.510$, compared to the corresponding LSUB$\infty(2)$
value $\delta_{c}^{>}=1.902$, for the case $\kappa=0$.  The true value
for $\delta_{c}^{>}$ at $\kappa=0$ certainly lies between these two
estimates.  A more detailed analysis of all our results yields our
best estimate as $\delta_{c}^{>} \approx 1.70(5)$.  Part of the
difference clearly comes from the fact that Eq.\ (\ref{Eq_exponFit})
retains only the leading-order correction to the large-$n$ limit,
while Eq.\ (\ref{M_extrapo_frustrated}) also retains both the leading and sub-leading corrections.  Nevertheless, it is likely that the scheme of Eq.\ (\ref{M_extrapo_frustrated}) does somewhat overestimate the
critical value $\delta_{c}^{>}$ in this case $\kappa=0$ when
frustration is absent.  We note that this level of discrepancy between
the two different extrapolation schemes diminishes rapidly as
intralayer frustration is introduced, such that for values
$\kappa \gtrsim 0.1$ it seems to be negligible.

In Figs.\ \ref{M_raw_extrapo_fix-J2}(b) and
\ref{M_raw_extrapo_fix-J2}(c) we also show comparable results to those
for the unfrustrated $(\kappa=0)$ case shown in Fig.\
\ref{M_raw_extrapo_fix-J2}(a), for the two cases where the intralayer
frustration parameter, $\kappa \equiv J_{2}/J_{1}$, takes the values
$\kappa=0.1$ and $\kappa=0.2$, respectively.  As expected, the effect
of increasing frustration is observed generally to reduce the N\'{e}el
order parameter $M$ at any given value of the interlayer coupling
strength $\delta$.  Accordingly, the upper critical value,
$\delta_{c}^{>}(\kappa)$, of the interlayer coupling strength, above
which N\'{e}el order vanishes, is seen to decrease monotonically as
$\kappa$ is increased.  Exactly as for the case $\kappa=0$ shown in
Fig.\ \ref{M_raw_extrapo_fix-J2}(a), the two LSUB$\infty$
extrapolations for $M$ based on the scheme of Eq.\
(\ref{M_extrapo_frustrated}), but with the two different LSUB$n$ data
sets with $n=\{2,6,10\}$ and $n=\{4,6,8,10\}$, agree extremely closely
with one another for all values of $\kappa$, with the agreement
generally even improving as $\kappa$ is increased.  For example, the
values for $\delta_{c}^{>}(\kappa)$ at the values $\kappa=0.1$ and
$\kappa=0.2$ shown in Figs.\ \ref{M_raw_extrapo_fix-J2}(b) and
\ref{M_raw_extrapo_fix-J2}(c) obtained from the two separate
LSUB$\infty$ extrapolations are $\delta_{c}^{>}(0.1)=1.275$,
$\delta_{c}^{>}(0.2)=0.538$ based on the LSUB$n$ data set with
$n=\{2,6,10\}$, and $\delta_{c}^{>}(0.1)=1.265$,
$\delta_{c}^{>}(0.2)=0.535$ based on the respective set
with$n=\{4,6,8,10\}$.

\begin{figure}[b]
  \includegraphics[angle=270,width=9cm]{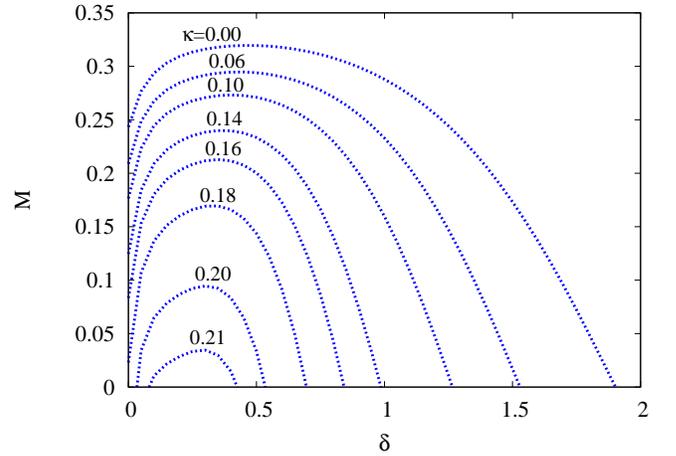}
  \caption{CCM results for the GS magnetic order parameter $M$ versus
    the scaled interlayer exchange coupling constant,
    $\delta \equiv J_{1}^{\perp}/J_{1}$, for the spin-$\frac{1}{2}$
    $J_{1}$--$J_{2}$--$J_{1}^{\perp}$ model on the bilayer honeycomb
    lattice (with $J_{1}>0$), for a variety of values of the
    intralayer frustration parameter, $\kappa \equiv J_{2}/J_{1}$.
  In each case we show extrapolated results, based on the N\'{e}el
  state as CCM model state, obtained from using Eq.\
  (\ref{M_extrapo_frustrated}) with the corresponding LSUB$n$ data
  sets $n=\{4,6,8,10\}$.}
\label{M_J2fix-selective_extrapo}
\end{figure}

\begin{figure*}[t]
\begin{center}
\mbox{
\hspace{-1.0cm}
\subfigure[]{\scalebox{0.3}{\includegraphics[angle=270]{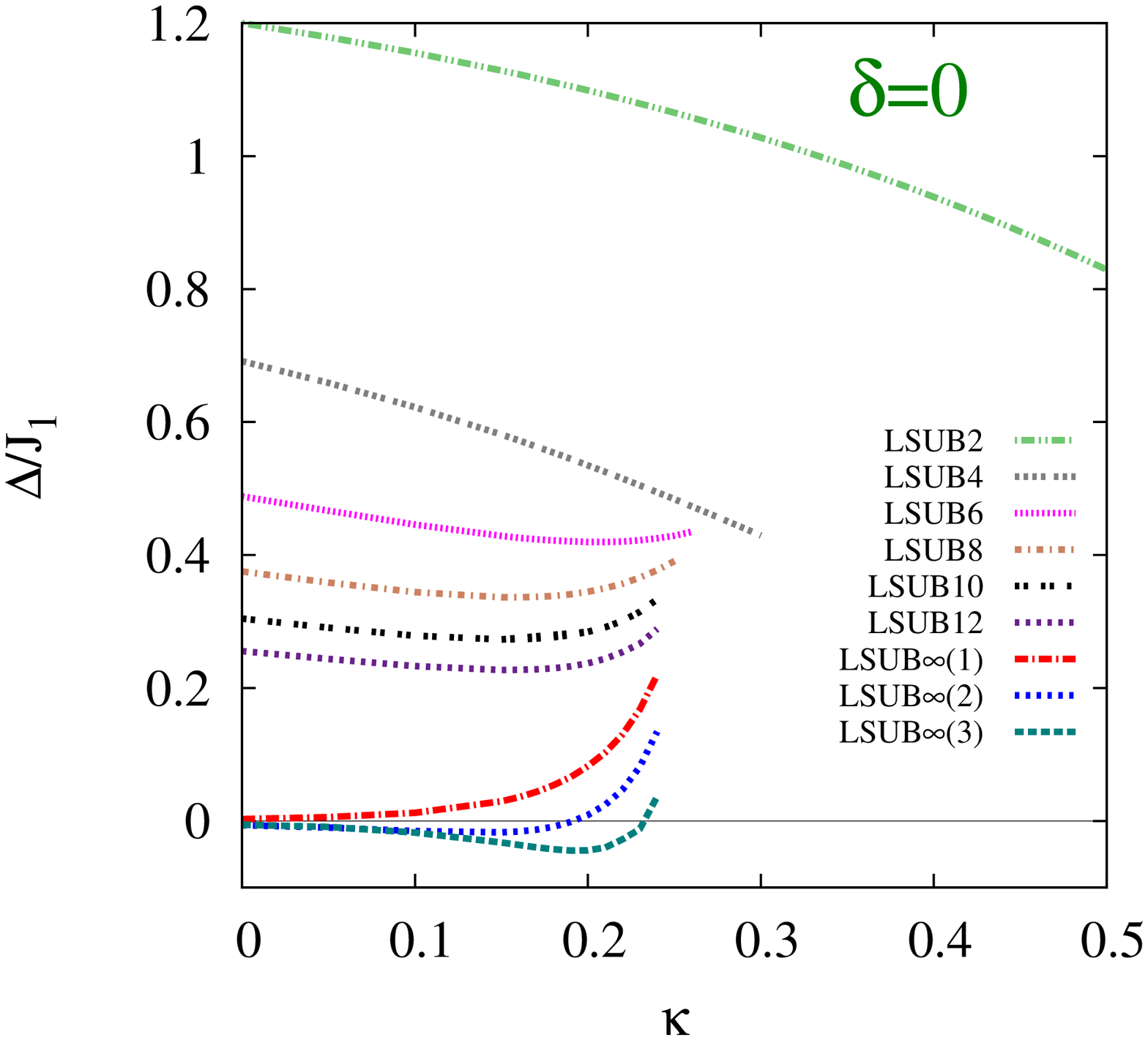}}}
\hspace{-1.8cm}
\subfigure[]{\scalebox{0.3}{\includegraphics[angle=270]{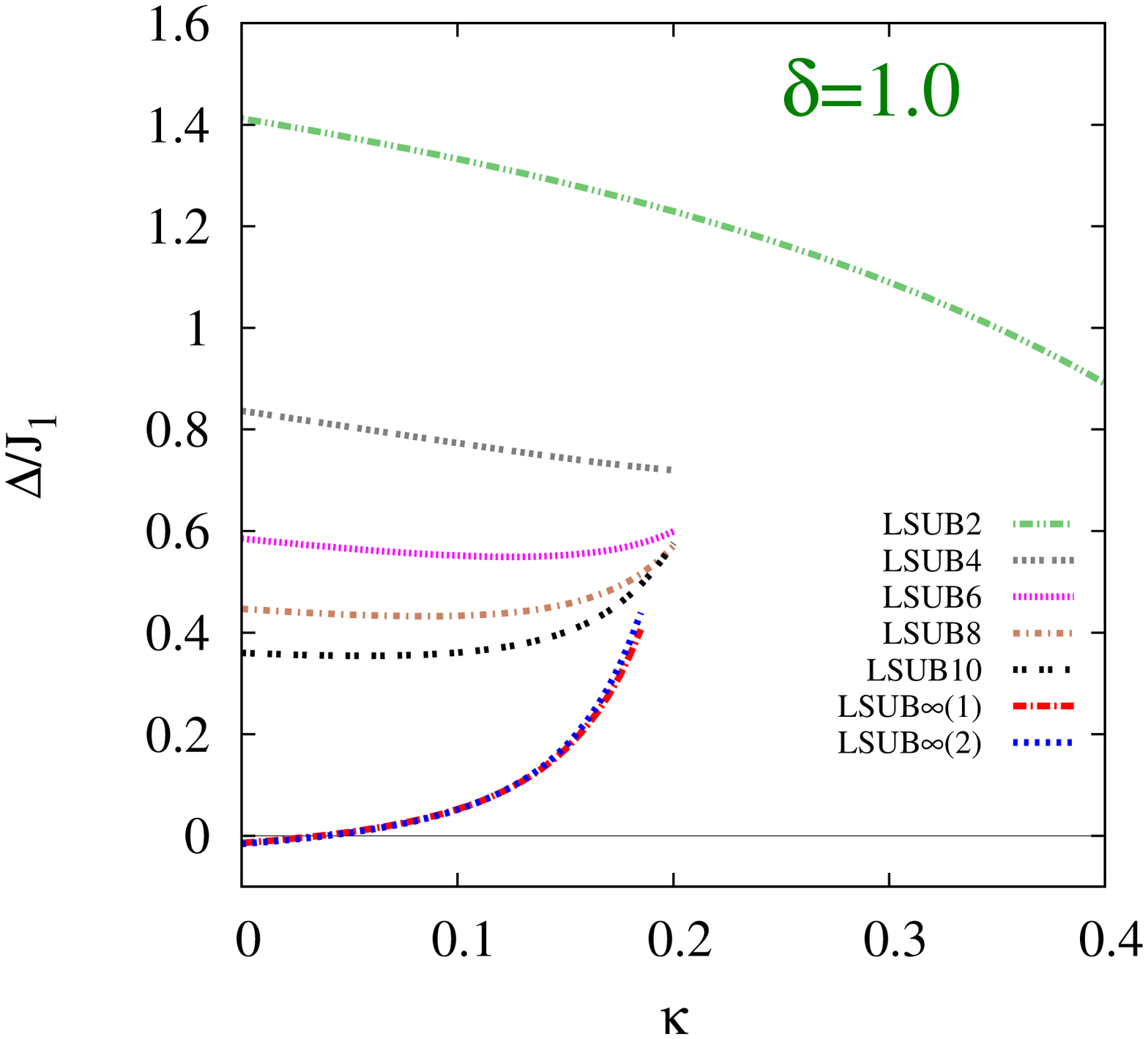}}}
\hspace{-1.8cm}
\subfigure[]{\scalebox{0.3}{\includegraphics[angle=270]{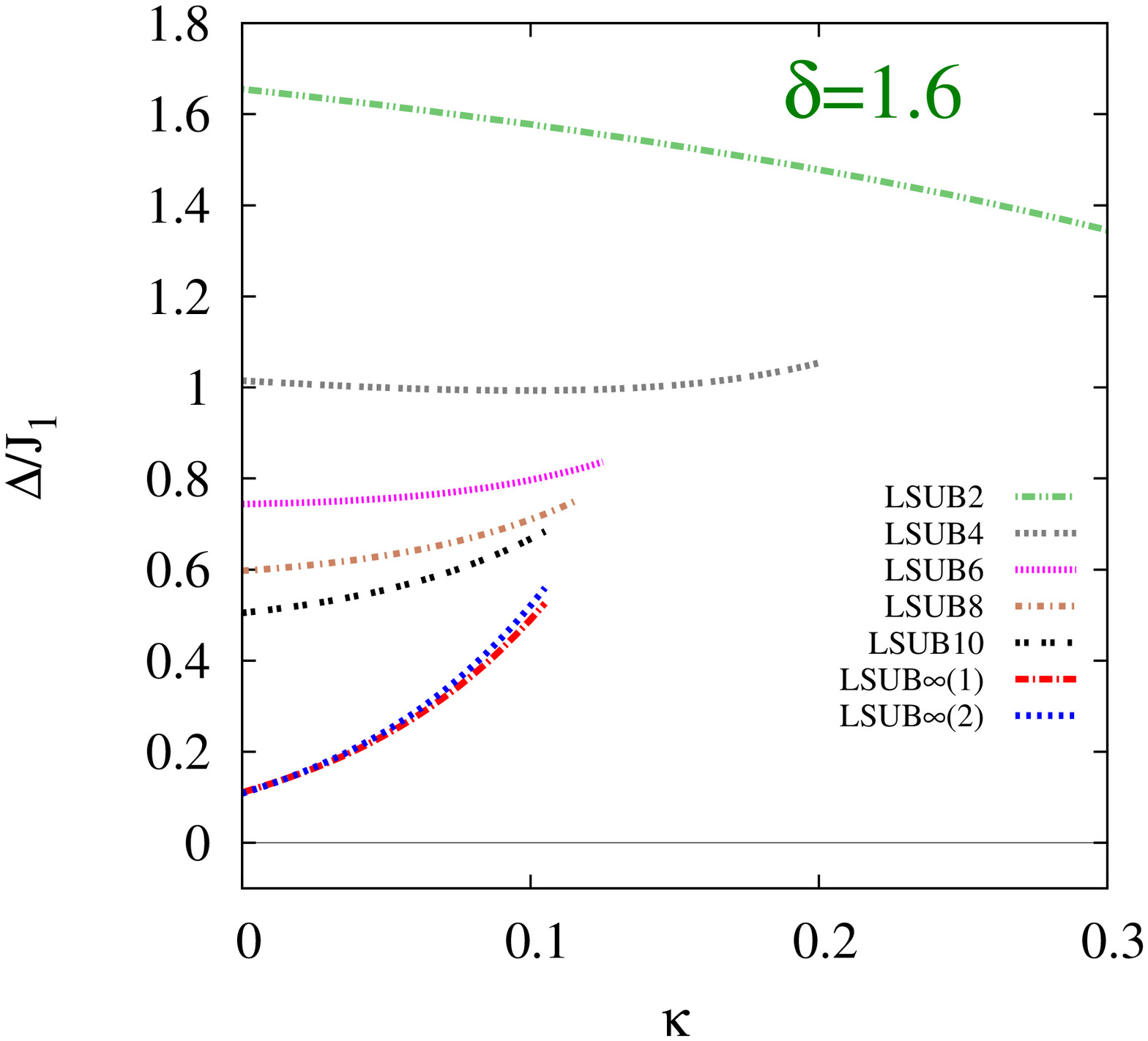}}}
}
\caption{CCM results for the triplet spin gap $\Delta$ (in units of $J_{1}$) versus
  the intralayer frustration parameter,
  $\kappa \equiv J_{2}/J_{1}$, for the spin-$\frac{1}{2}$
  $J_{1}$--$J_{2}$--$J_{1}^{\perp}$ model on the bilayer honeycomb
  lattice (with $J_{1}>0$), for three selected values of the scaled
  interlayer exchange coupling constant, $\delta \equiv J_{1}^{\perp}/J_{1}$:
  (a) $\delta=0$, (b) $\delta=1.0$, and (c) $\delta=1.6$.  Results
  based on the N\'{e}el state as CCM model state are shown in LSUB$n$
  approximations with $n=2,4,6,8,10$ (and also with $n=12$ for the
  special case of the $J_{1}$--$J_{2}$ monolayer, i.e., when
  $\delta=0$), together with various corresponding LSUB$\infty(i)$
  extrapolated results using Eq.\ (\ref{Eq_spin_gap}) and the
  respective data sets $n=\{2,6,10\}$ for $i=1$, $n=\{4,6,8,10\}$ for
  $i=2$, and $n=\{4,8,12\}$ for $i=3$ (for the case $\delta=0$ only).}
\label{Egap_raw_extrapo_fix-J1perp}
\end{center}
\end{figure*}  

In Fig.\ \ref{M_raw_extrapo_fix-J2}(c) the value $\kappa=0.2$ for
which the order parameter $M$ is shown as a function of $\delta$ is
greater than the above-quoted values $\kappa_{c}(0)$ above which
N\'{e}el order vanishes in the monolayer $(\delta=0)$, for both
LSUB$\infty$ extrapolations displayed.  Interestingly now, however, for
values $\kappa > \kappa_{c}(0)$ that are not too large, as $\delta$ is increased above a
lower critical value $\delta_{c}^{<}(\kappa)$, N\'{e}el order is
re-established due to the interlayer AFM coupling, up to the
respective upper critical value $\delta_{c}^{>}(\kappa)$.  For the
value $\kappa=0.2$ shown in Fig.\ \ref{M_raw_extrapo_fix-J2}(c), for
example, we find that the values for $\delta_{c}^{<}(0.2)$ obtained from
the two separate LSUB$\infty$ extrapolations shown are
$\delta_{c}^{<}(0.2)=0.041$ based on the LSUB$n$ data set with
$n=\{2,6,10\}$, and $\delta_{c}^{<}(0.2)=0.033$ based on that
with $n=\{4,6,8,10\}$.  This behavior is further illustrated in Fig.\
\ref{M_J2fix-selective_extrapo}, from which we clearly observe that
there exists some upper critical value $\kappa^{>}$ of the intralayer
frustration parameter $\kappa$ at which
$\delta_{c}^{<}(\kappa^{>})=\delta_{c}^{>}(\kappa^{>})$, and hence
such that N\'{e}el order is absent for all values
$\kappa > \kappa^{>}$, whatever the value of $\delta$.
Our corresponding LSUB$\infty$ estimates for $\kappa^{>}$ are
$\kappa^{>}=0.216$ based on the LSUB$n$ data set with $n=\{2,6,10\}$ and
$\kappa^{>}=0.214$ based on the set with $n=\{4,6,8,10\}$.

\begin{figure*}[t]
\begin{center}
\mbox{
\hspace{-1.0cm}
\subfigure[]{\scalebox{0.3}{\includegraphics[angle=270]{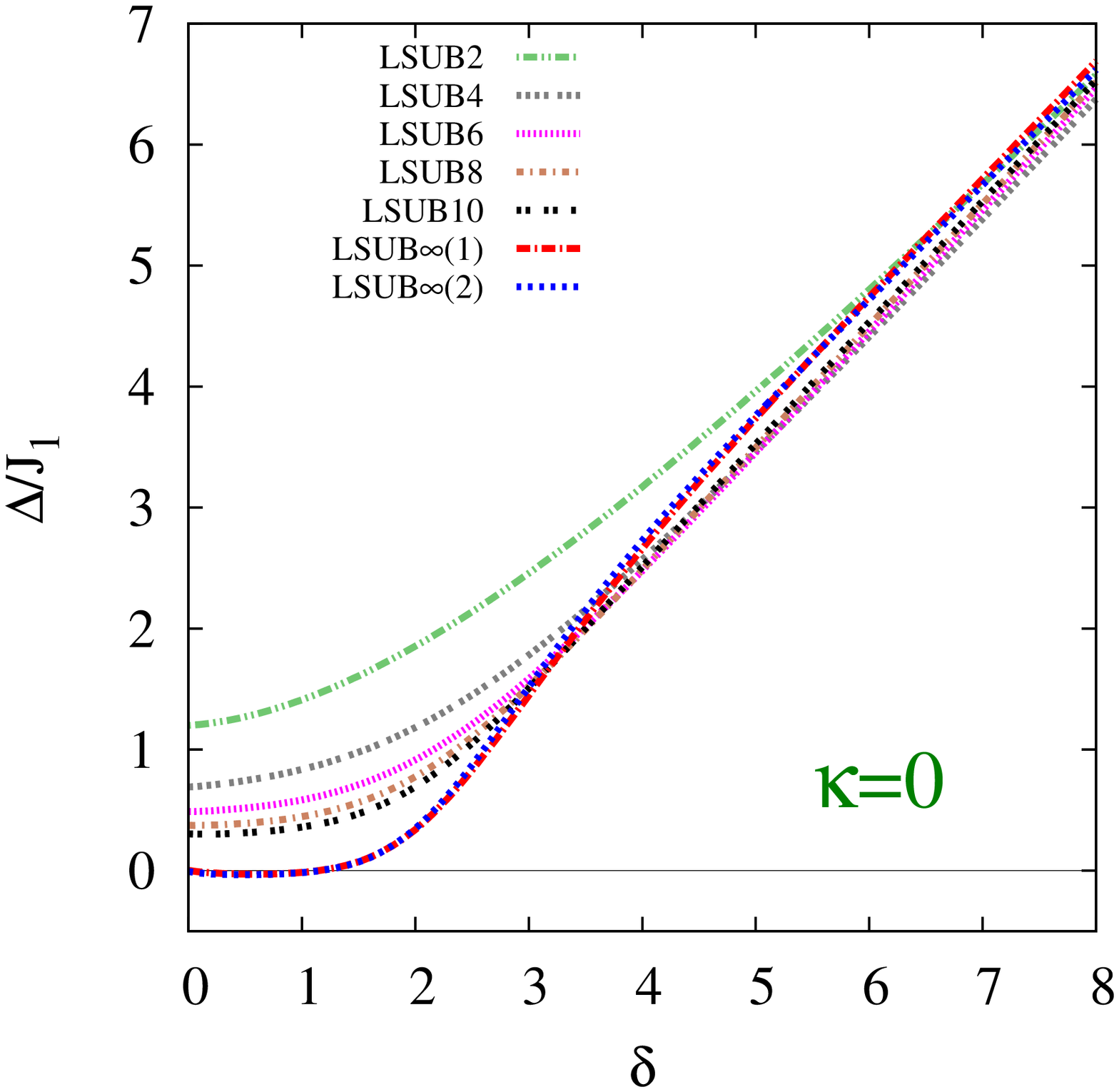}}}
\hspace{-1.8cm}
\subfigure[]{\scalebox{0.3}{\includegraphics[angle=270]{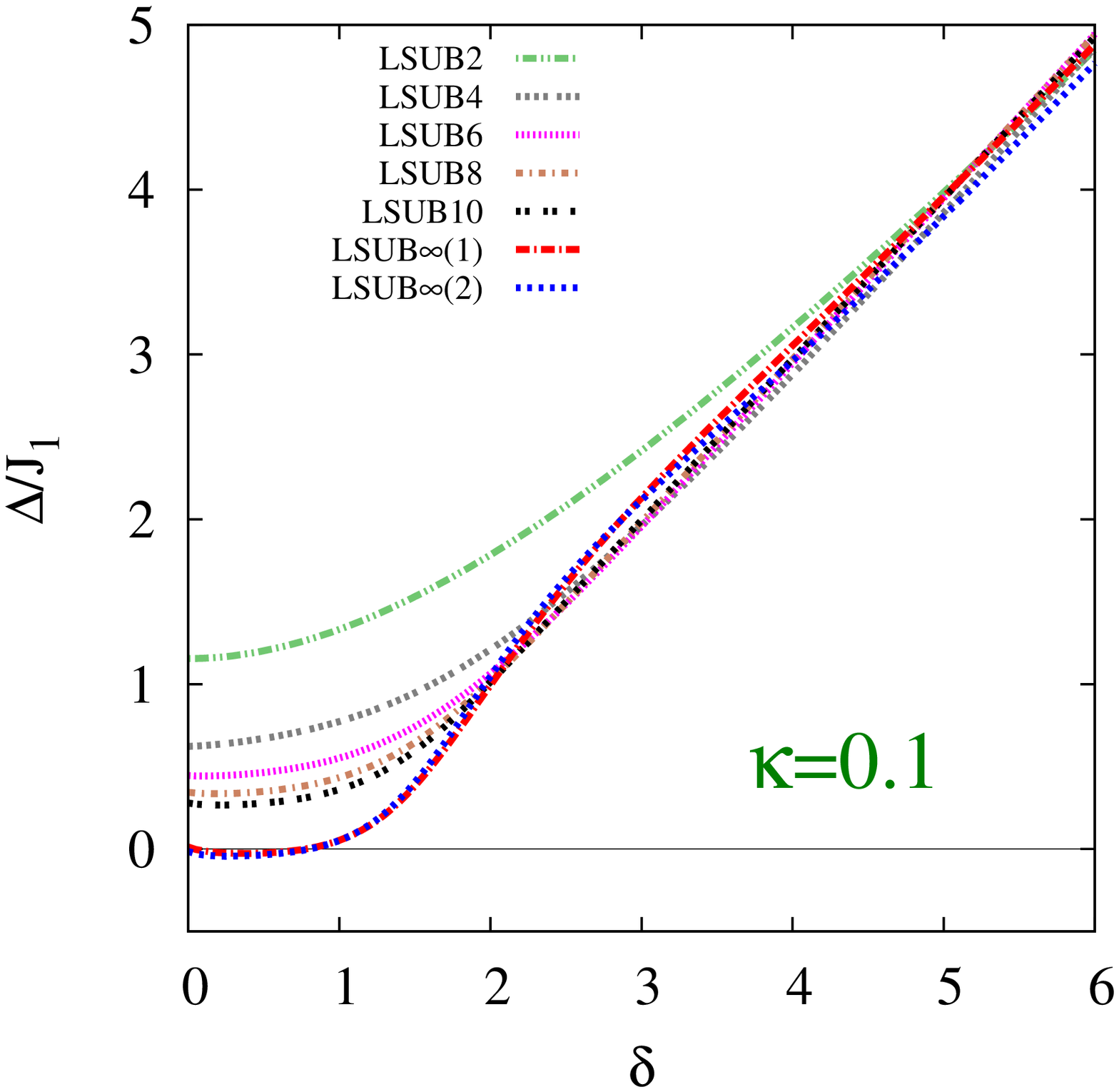}}}
\hspace{-1.8cm}
\subfigure[]{\scalebox{0.3}{\includegraphics[angle=270]{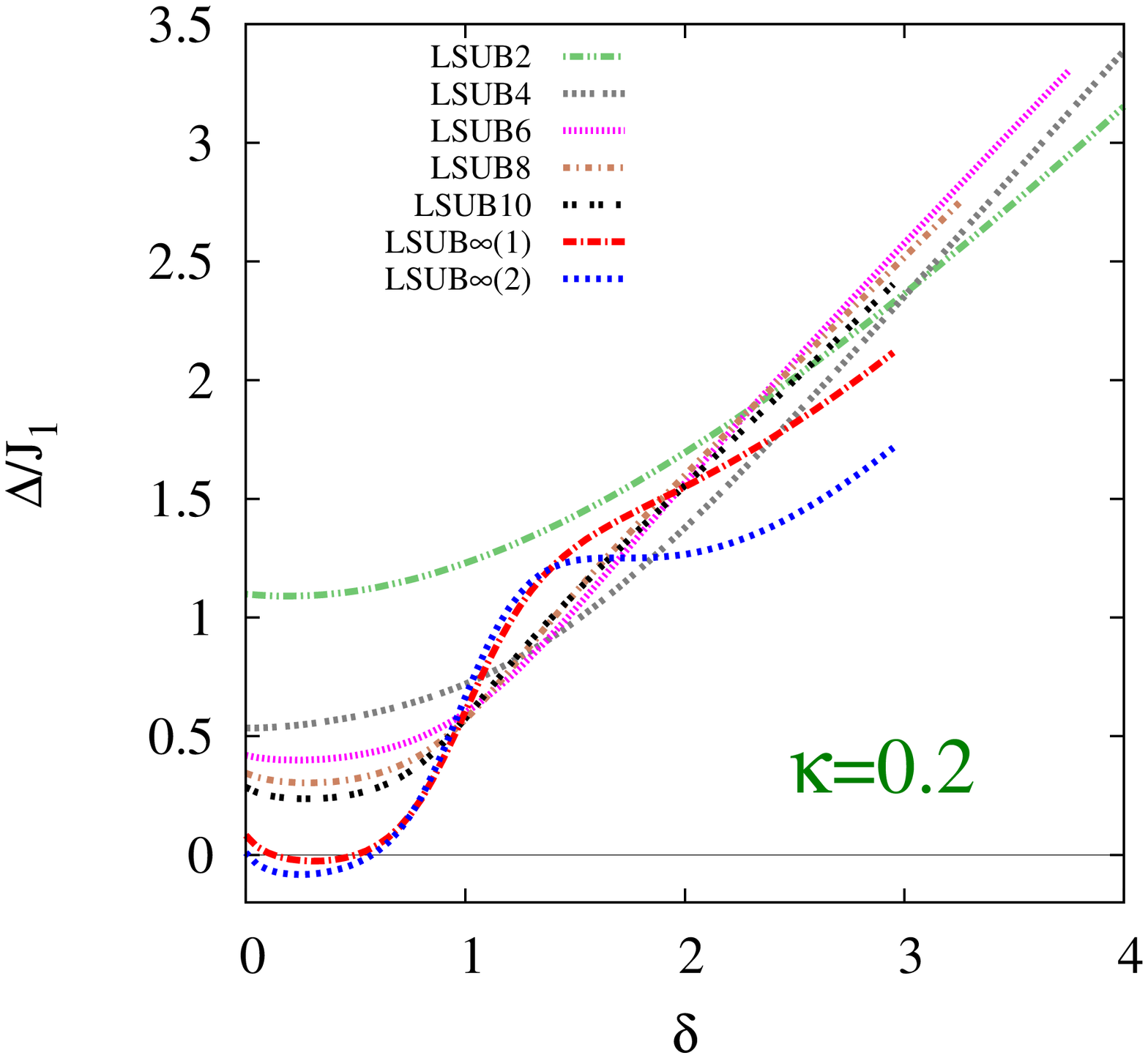}}}
}
\caption{CCM results for the triplet spin gap $\Delta$ (in units
  of $J_{1}$) versus the scaled interlayer exchange coupling constant,
  $\delta \equiv J_{1}^{\perp}/J_{1}$, for the spin-$\frac{1}{2}$
  $J_{1}$--$J_{2}$--$J_{1}^{\perp}$ model on the bilayer honeycomb
  lattice (with $J_{1}>0$), for three selected values of the
  intralayer frustration parameter, $\kappa \equiv J_{2}/J_{1}$: (a)
  $\kappa=0$, (b) $\kappa=0.1$, and (c) $\kappa=0.2$.  Results based
  on the N\'{e}el state as CCM model state are shown in LSUB$n$
  approximations with $n=2,4,6,8,10$, together with two corresponding
  LSUB$\infty(i)$ extrapolated results using Eq.\ (\ref{Eq_spin_gap})
  and the respective data sets $n=\{2,6,10\}$ for $i=1$ and
  $n=\{4,6,8,10\}$ for $i=2$.}
\label{Egap_raw_extrapo_fix-J2}
\end{center}
\end{figure*}

\begin{figure*}[t]
\begin{center}
\mbox{
\hspace{-0.6cm}
\subfigure[]{\scalebox{0.263}{\includegraphics[angle=270]{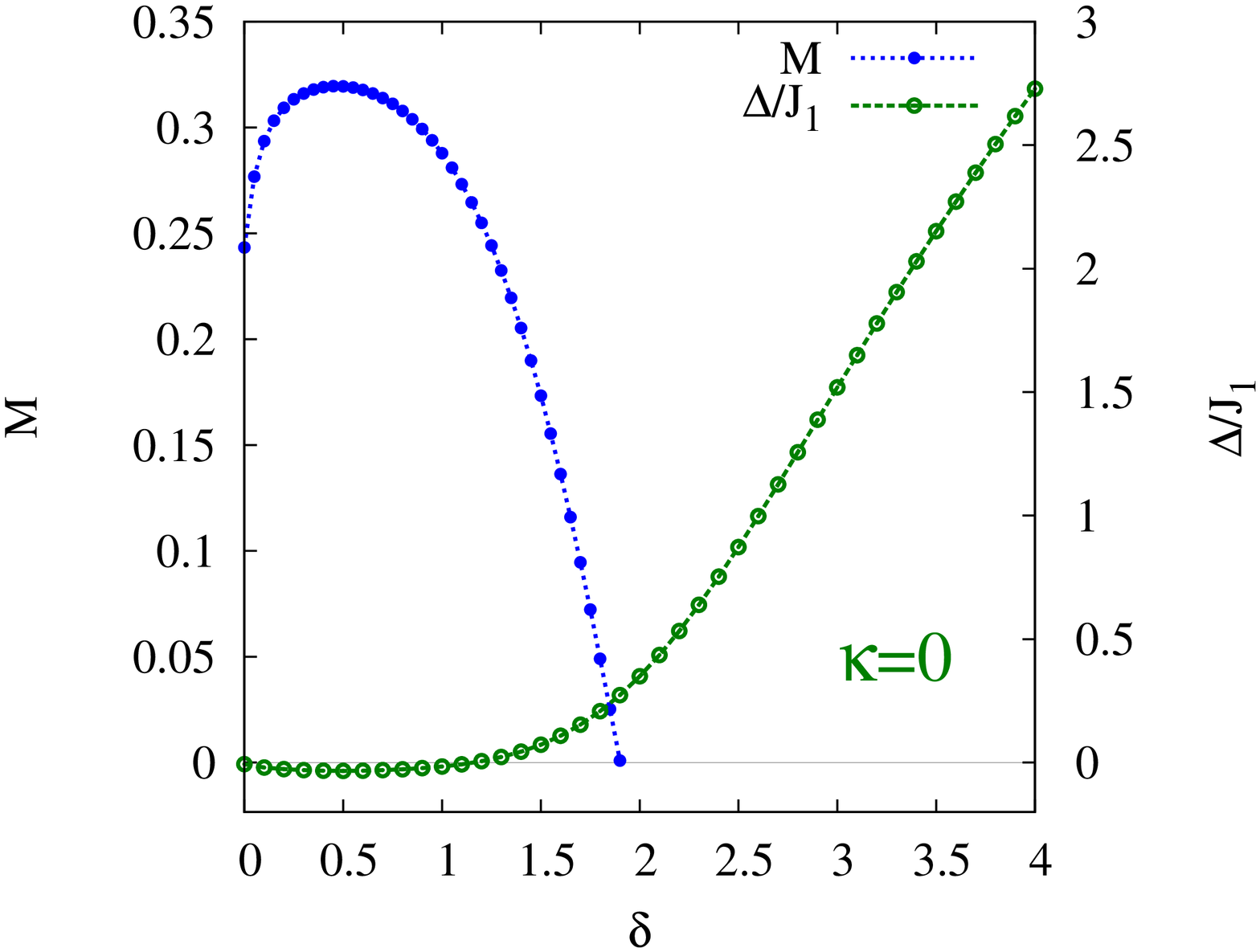}}}
\hspace{-0.9cm}
\subfigure[]{\scalebox{0.263}{\includegraphics[angle=270]{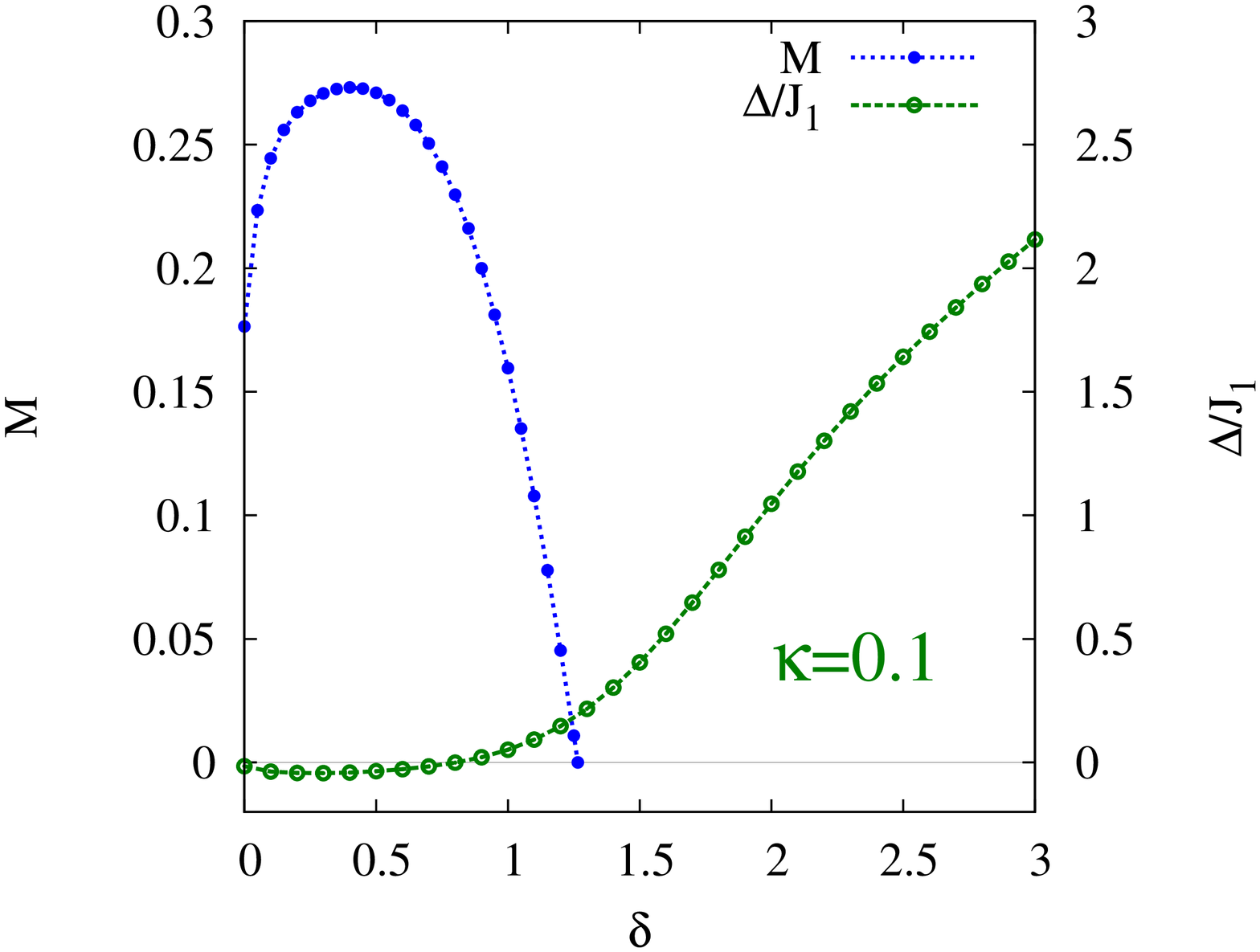}}}
\hspace{-0.9cm}
\subfigure[]{\scalebox{0.263}{\includegraphics[angle=270]{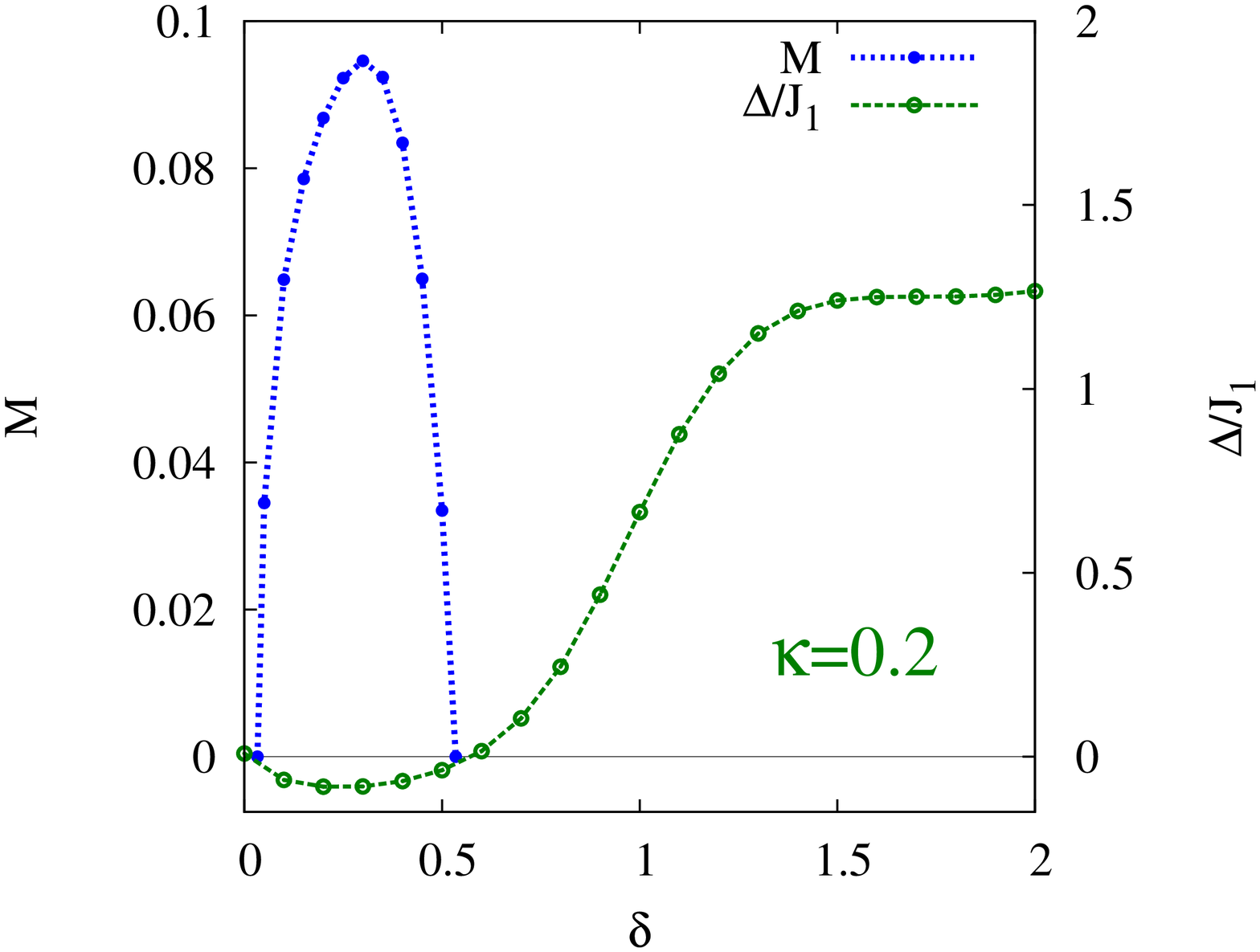}}}
}
\caption{Juxtaposed CCM results for the magnetic order parameter $M$ (left
  scale) and the triplet spin gap $\Delta$ (in units of $J_{1}$, right
  scale) versus the scaled interlayer exchange coupling constant,
  $\delta \equiv J_{1}^{\perp}/J_{1}$, for the spin-$\frac{1}{2}$
  $J_{1}$--$J_{2}$--$J_{1}^{\perp}$ model on the bilayer honeycomb
  lattice (with $J_{1}>0$), for three selected values of the
  intralayer frustration parameter, $\kappa \equiv J_{2}/J_{1}$: (a)
  $\kappa=0$, (b) $\kappa=0.1$, and (c) $\kappa=0.2$.  Extrapolated
  results for $M$ and $\Delta$ are shown from using Eqs.\
  (\ref{M_extrapo_frustrated}) and (\ref{Eq_spin_gap}), respectively,
  with the corresponding LSUB$n$ data sets with $n=\{4,6,8,10\}$ in
  each case, based on the N\'{e}el state as the CCM model state.}
\label{M_Egap_extrapo_fix-J2}
\end{center}
\end{figure*}  

\begin{figure}[!t]
  \includegraphics[angle=270,width=9cm]{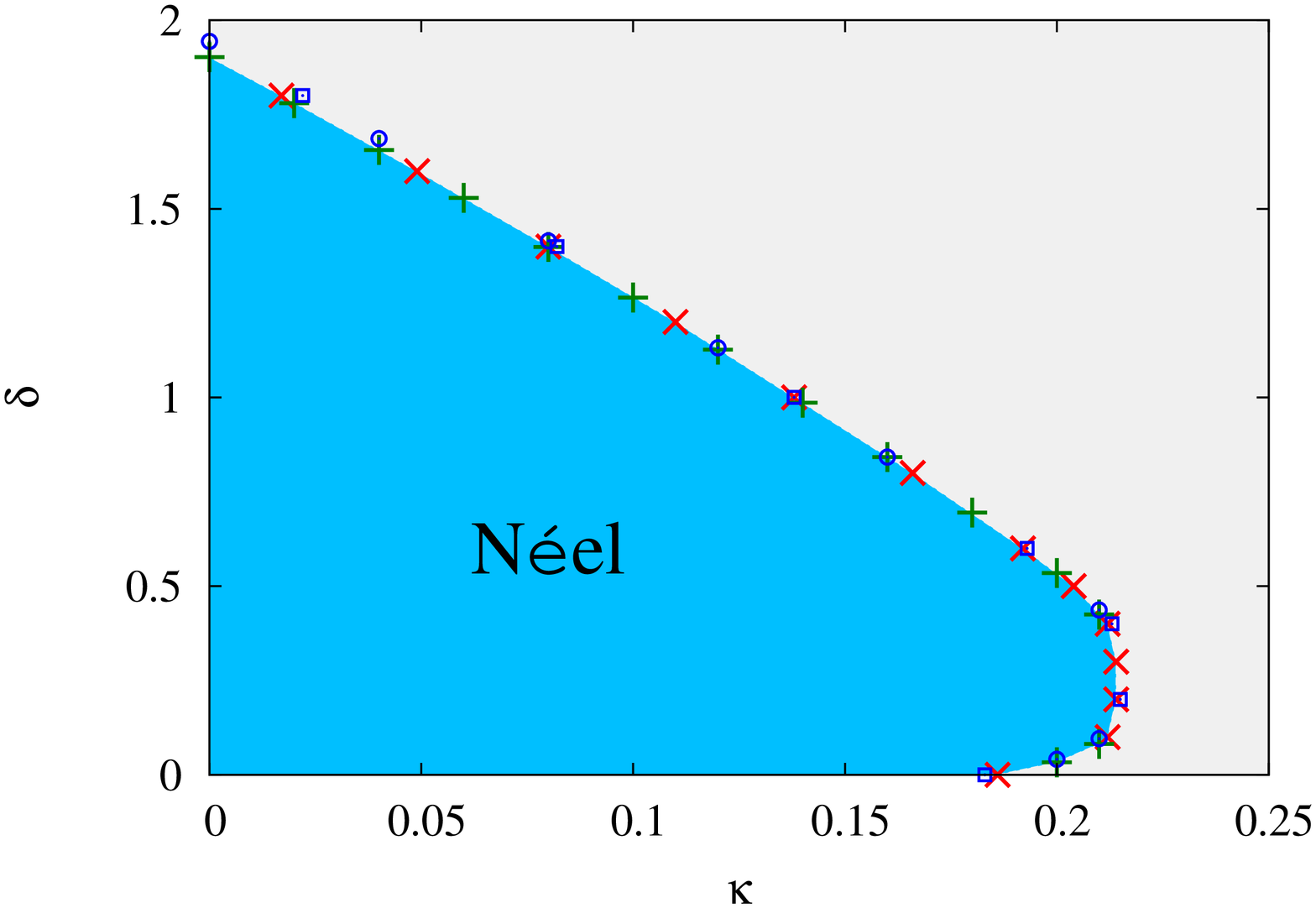}
  \caption{$T=0$ phase diagram of the spin-$\frac{1}{2}$
    $J_{1}$--$J_{2}$--$J_{1}^{\perp}$ model on the bilayer honeycomb
    lattice with $J_{1}>0$, $\delta \equiv J_{1}^{\perp}/J_{1}$, and
    $\kappa \equiv J_{2}/J_{1}$.  The darker (skyblue) region is the
    quasiclassical phase with AFM N\'{e}el order, while in the lighter
    (grey) region N\'{e}el order is absent.  The red cross ($\times$)
    symbols and the green plus ($+$) symbols are points at which the
    extrapolated GS magnetic order parameter $M$ for the N\'{e}el
    phase vanishes, for specified values of $\delta$ and $\kappa$,
    respectively.  They thus represent the values $\kappa_{c}(\delta)$
    and $\delta_{c}^{>}(\kappa)$ [and also $\delta_{c}^{<}(\kappa)$
    for values of $\kappa$ in the range
    $\kappa_{c}(0) < \kappa < \kappa^{>}$], respectively.  In each
    case the N\'{e}el state is used as CCM model state, and Eq.\
    (\ref{M_extrapo_frustrated}) is used for the extrapolations with
    the corresponding LSUB$n$ data sets $n=\{4,6,8,10\}$.  For
    comparison we also show by blue open square ($\square$) and open
    circle ($\bigcirc$) symbols respectively, some corresponding
    points obtained using the LSUB$n$ data set $n=\{2,6,10\}$.}
\label{phase_diag}
\end{figure}

We turn next to our corresponding results for the triplet spin gap
$\Delta$.  Firstly, we show in Fig.\ \ref{Egap_raw_extrapo_fix-J1perp}
corresponding sets of results as functions of $\kappa$ to those shown
in Fig.\ \ref{M_raw_extrapo_fix-J1perp} for the N\'{e}el magnetic
order parameter $M$, for the same three fixed values of $\delta$.
Figure \ref{Egap_raw_extrapo_fix-J1perp}(a) shows our LSUB$n$ results
for the spin-$\frac{1}{2}$ honeycomb-lattice monolayer (i.e., with
$\delta=0$), where again for this limiting case alone calculations are
presented for values $n \leq 12$ of the LSUB$n$ truncation parameter.
As before, in Fig.\ \ref{M_raw_extrapo_fix-J1perp}(a) for $M$, we now
also show three separate LSUB$\infty$ extrapolations, based
respectively on the LSUB$n$ data sets $n=\{2,6,10\}$,
$n=\{4,6,8,10\}$, and $n=\{4,8,12\}$, and using the scheme of Eq.\
(\ref{Eq_spin_gap}) in each case.  It is evident that each of the
extrapolations is consistent, within small numerical errors, with the
gap being zero in the region $\kappa < \kappa_{c}(0)$ where N\'{e}el
LRO is present, exactly as expected.  There is also clear evidence
that for values $\kappa > \kappa_{c}(0)$ the GS phase is gapped,
consistent with it being a VBC state.  Corresponding results for
$\Delta = \Delta(\kappa)$ are shown in Fig.\
\ref{Egap_raw_extrapo_fix-J1perp}(b) and
\ref{Egap_raw_extrapo_fix-J1perp}(c) for the bilayer with values
$\delta = 1.0$ and $\delta = 1.6$ respectively of the interlayer
coupling strength.

Once again, the effect of the interlayer coupling on the spin gap
$\Delta$ is also displayed separately in Fig.\
\ref{Egap_raw_extrapo_fix-J2} for the same three different values of
the intralayer frustration parameter $\kappa$ as are shown in Fig.\
\ref{M_raw_extrapo_fix-J2} for the N\'{e}el order parameter $M$.  Figure \ref{Egap_raw_extrapo_fix-J2}(a) in particular clearly shows that $\Delta/J_{1} \rightarrow \delta$ in the large $\delta$ limit, exactly as expected for the IDVBC state, from Eq.\ (\ref{IDVBC_triplet_spin_gap_limit-J1perp-infty_eq}).  Each
of the LSUB$\infty$ extrapolations shown gives results for the system
being gapless over ranges of values of $\kappa$ that are in striking
agreement with the ranges shown by the corresponding LSUB$\infty$
extrapolations in Fig.\ \ref{M_raw_extrapo_fix-J2} for where N\'{e}el
LRO survives.  For the specific case $\kappa=0.2$ shown in Fig.\
\ref{Egap_raw_extrapo_fix-J2}(c), for example, there is clear evidence
of both a lower critical value $\delta_{c}^{<}(0.2)$ and an upper
critical value $\delta_{c}^{>}(0.2)$, between which the stable GS
phase is gapless, and the values so obtained from the gap $\Delta$ are
in excellent agreement with those values shown in Fig.\
\ref{M_raw_extrapo_fix-J2}(c) between which the N\'{e}el order
parameter $M$ is nonzero.  The overall level of agreement is shown
even more clearly in Fig.\ \ref{M_Egap_extrapo_fix-J2}, where our
extrapolated CCM results for $M$ and $\Delta$ are juxtaposed on the
same graph for the same three specific cases $\kappa=0$, $\kappa=0.1$,
and $\kappa=0.2$.  We note that the only significant disagreement
between the values $\delta_{c}^{>}(\kappa)$ obtained from the
extrapolations based on $M$ and $\Delta$ arises at $\kappa=0$.  Here,
as we have discussed previously, and shown in Fig.\
\ref{M_raw_extrapo_fix-J2}(a), it is likely that the extrapolation for
$M$ based on Eq.\ (\ref{M_extrapo_frustrated}) slightly overestimates
the value $\delta_{c}^{>}(0)$.  Correcting for this as we have
indicated would thereby bring it into better agreement with the value
obtained from $\Delta$.

We note that, within the CCM framework, the vanishing of the magnetic
order parameter $M$ generally provides an appreciably more accurate
estimate than the opening of a spin gap $\Delta$ for the critical
coupling that marks the transition from a gapless state with
quasiclassical magnetic LRO to a gapped non-magnetic state.  The
reason for this is that in most cases at the critical points where the
(extrapolated) order parameter $M$ vanishes the slope of the curve for
$M$ as a function of the respective coupling parameter is nonzero.
This behavior is clearly seen here in Figs.\
\ref{M_raw_extrapo_fix-J2} and \ref{M_J2fix-selective_extrapo}, for
example.  By contract, the extrapolated curves for the spin gap
$\Delta$ generally depart from being zero (at the same respective
critical points) with zero slope, as one sees here from
Figs. \ref{Egap_raw_extrapo_fix-J1perp} and
\ref{Egap_raw_extrapo_fix-J2}, for example.  Correspondingly, the CCM
estimates for the critical points have much larger errors than those
obtained from the vanishing of $M$.  What is interesting, however, is
that if one assumes that the exact gap disappears at a critical point,
as a function of the coupling, with non-zero (and possibly infinite) slope, a simple
extrapolation of the CCM results in Fig.\
\ref{Egap_raw_extrapo_fix-J2}, for example, using values somewhat
larger than the actual critical point (i.e., beyond where the curves
shown start appreciably, and thence presumably artificially, to round)
gives revised estimates for the critical points that are in remarkable
closer agreement to those obtained from the vanishing of $M$.  In this
context we note that for the specific case $\kappa=0$ of the model
with NN bonds only, the spin gap $\Delta$ has a singularity at the
critical value $\delta_{c}^{>}(0)$ with a critical exponent $\nu$
given by the three-dimensional Heisenberg universal value
$\nu \approx 0.71$ \cite{LeGuillou:1985_3d_Heiserberg}.  This exact
result lends credence to our assumption that the observed rounding of
our CCM results for $\Delta$ very near the critical points at which it
vanishes is itself an inherent error associated with the
extrapolation.

From our discussion above it is clear that the extrapolation of our
CCM LSUB$n$ data for the spin gap is more subtle than that for the
order parameter in a small region very close to a quantum critical
point.  Away from any such regions the errors are usually quite small,
however.  They can be quantified, for example, by comparing
LSUB$\infty$ extrapolations based on different LSUB$n$ input data
sets.  Any differences are an indicator of the size of the associated
errors.  Figure \ref{Egap_raw_extrapo_fix-J2} gives some typical
comparisons, and shows how robust our results are, in general.  As
noted already in Sec.\ \ref{ccm_sec}, extrapolations based on more
LSUB$n$ data points than fitting parameters are, as expected, more
accurate than those based on an equal number.  Figure
\ref{Egap_raw_extrapo_fix-J1perp}(a) demonstrates this point
particularly clearly.  A further internal check on the accuracy of the
LSUB$\infty$ extrapolations for $\Delta$ is the size of any observed
deviations from a zero value in the N\'{e}el-ordered regimes where the
system is certainly gapless (i.e., with soft Goldstone modes).  The
results in Fig.\ \ref{Egap_raw_extrapo_fix-J2} show very clearly that
the deviations from zero are very small, typically no greater than
$\pm 0.04$ for the relevant quantity $\Delta/J_{1}$.  For a
fit to the three-parameter scheme of Eq.\ (\ref{Eq_spin_gap}) using
four or more LSUB$n$ data points, such as the LSUB$\infty(2)$
extrapolation in Fig.\ \ref{Egap_raw_extrapo_fix-J2}, we may also
calculate the least-squares errors associated with the fit.  In
essentially all cases these error bars are entirely consistent with
$\Delta$ being zero in the expected regions
$\delta_{c}^{<}(\kappa) < \delta < \delta_{c}^{>}(\kappa)$.

Finally, in Fig.\ \ref{phase_diag}, we show our results for the $T=0$
phase diagram of the model in the $\kappa\delta$ plane.
The darker shaded area represents the region in which the stable GS
phase has N\'{e}el LRO as determined by the non-vanishing of our
LSUB$\infty$ estimates for the corresponding magnetic order parameter
$M$, obtained from extrapolating the LSUB$n$ results with Eq.\
(\ref{M_extrapo_frustrated}) and using the data sets $n=\{4,6,8,10\}$
as input.  We show by different symbols results obtained for
$\kappa_{c}(\delta)$ at fixed values of $\delta$ from curves such as
those shown in Fig.\ \ref{M_raw_extrapo_fix-J1perp}, and for
$\delta_{c}^{>}(\kappa)$ at fixed values of $\kappa$ [and also
$\delta_{c}^{<}(\kappa)$ for values of $\kappa$ in the range
$\kappa_{c}(0) < \kappa < \kappa^{>}$] from curves such as those shown
in Figs.\ \ref{M_raw_extrapo_fix-J2} and
\ref{M_J2fix-selective_extrapo}.  The fact that the corresponding
points on the phase boundary from two different sets of results agree
so well with each other is testament to the accuracy of the
extrapolation scheme.  To indicate the relative insensitivity of our
prediction for the N\'{e}el phase boundary, we also show in Fig.\
\ref{phase_diag} some selected corresponding points using the LSUB$n$
data points with $n=\{2,6,10\}$ for the extrapolations.  Clearly, the
results are robust.

We summarize and discuss the results in Sec.\
\ref{discuss_summary_sec}, where we also compare them with the other
limited results available.

\section{DISCUSSION AND SUMMARY}
\label{discuss_summary_sec}
In this paper we have investigated the $T=0$ quantum phase diagram of
the frustrated spin-$\frac{1}{2}$ $J_{1}$--$J_{2}$--$J_{1}^{\perp}$
model on the honeycomb bilayer lattice.  To that end we have used the
fully systematic approach afforded by the CCM, which we have
implemented computationally to very high orders in the well-defined
LSUB$n$ hierarchy of approximations.  The method has the particular
unique advantages that it exactly obeys both the Goldstone
linked-cluster theorem and the Hellmann-Feynman theorem at all levels
of approximation.  The former property guarantees that the method is
fully size-extensive.  Hence, all our calculations are performed from
the outset in the thermodynamic limit of an infinite lattice
$(N \rightarrow \infty)$, thereby obviating the need for any
finite-size scaling of the results, as is required in most alternative
high-precision methods.  Furthermore, no other approximations are
made, apart from the choice of LSUB$n$ truncation-order parameter $n$.
We have performed calculations up to order $n=10$, and our sole source
of error lies in extrapolating our LSUB$n$ sequences of results for
the physical parameters calculated to the exact
$(n \rightarrow \infty)$ limit.  Such series of calculations (for
$n \leq 10$) and extrapolations $(n \rightarrow \infty)$ have been
made for each of the GS energy per spin, the GS N\'{e}el magnetic
order parameter (i.e., the staggered magnetization), and the energy
gap to the lowest-lying excited spin-triplet state.

From such results we have accurately determined the phase boundary,
$\kappa = \kappa_{c}(\delta)$ or, equivalently,
$\delta=\delta_{c}(\kappa)$, in the $\kappa\delta$ plane (where
$\kappa \equiv J_{2}/J_{1}$ is the intralayer frustration parameter,
and $\delta \equiv J_{1}^{\perp}/J_{1}$ is the interlayer coupling
parameter) on which N\'{e}el AFM order melts and inside which the
system has N\'{e}el magnetic LRO.  Our main findings can be summarized
as follows:
\begin{itemize}
\item For all values $\delta < \delta_{c}^{>}(0) \approx 1.70(5)$
  there exists an upper critical value $\kappa_{c}(\delta)$, so that
  the system has N\'{e}el LRO for $\kappa < \kappa_{c}(\delta)$.
\item For all values $\kappa < \kappa_{c}(0) \approx 0.19(1)$ there
  exists an upper critical value $\delta_{c}^{>}(\kappa)$, so that the
  system has N\'{e}el LRO for $0 < \delta < \delta_{c}^{>}(\kappa)$.
\item For slightly higher values of $\kappa$ in the range
  $\kappa_{c}(0) < \kappa < \kappa^{>} \approx 0.215(2)$ there exists
  a reentrant region in the phase diagram such that the system has
  N\'{e}el LRO only in the range
  $\delta_{c}^{<}(\kappa) < \delta < \delta_{c}^{>}(\kappa)$, where $\delta_{c}^{<}(\kappa)>0$.
\item The lower and upper critical values, $\delta_{c}^{<}(\kappa)$
  and $\delta_{c}^{>}(\kappa)$, respectively, coalesce when
  $\kappa = \kappa^{>}$, such that
  $\delta_{c}^{<}(\kappa^{>}) = \delta_{c}^{>}(\kappa^{>}) \approx
  0.25(5)$.
\item N\'{e}el order is absent outside the region defined above.
 \end{itemize}

 The reentrant behavior itself finds a simple explanation, as we
 discuss below.  The $J_{1}$--$J_{2}$ honeycomb-lattice monolayer
 (i.e., $\delta=0$) has perfect N\'{e}el magnetic LRO at the classical
 ($s \rightarrow \infty$) level for
 $\kappa < \kappa_{{\rm cl}}(0) \equiv \frac{1}{6}$.  At this
 classical critical point there is a phase transition to a state with
 spiral order.  As one expects, the effects of quantum fluctuations
 are to preserve the collinearly ordered state to somewhat higher
 values of the frustration parameter $\kappa$ beyond which the system develops
 noncollinear order.  For the spin-$\frac{1}{2}$ system we have found
 the corresponding value $\kappa_{c}(0) \approx 0.19(1)$.  Near this
 critical point the addition of a small positive interlayer coupling
 $\delta$ should further enhance the stability of the N\'{e}el phase,
 as a small value of the bilayer coupling makes the system {\it more}
 ordered.  Naturally, if one increases the bilayer coupling too much
 the N\'{e}el order will start to compete with the formation of IDVBC
 order.  Thus, the reentrant behavior found in the phase diagram for
 values of $\kappa > \kappa_{c}(0)$ is not surprising.  What is more
 surprising perhaps is its very limited extent.

 Finally, it is of interest to compare our results with those using
 other high-precision methods in the few limiting cases that are
 available in the literature.  For example, only in the limiting case
 $\kappa=0$ of no intralayer frustration is every lattice bond a link
 between two sites belonging to different sublattices of the bipartite
 bilayer honeycomb lattice.  For this limiting case alone no ``sign
 problem'' arises, and the unfrustrated $J_{1}$--$J_{1}^{\perp}$ model
 is amenable to QMC simulation.  Such a QMC simulation of the
 spin-$\frac{1}{2}$ $J_{1}$--$J_{1}^{\perp}$ honeycomb bilayer model
 has been performed \cite{Ganesh:2011_honey_bilayer_PRB84} on lattices
 with linear size $L \leq 36$, using the stochastic series expansion
 algorithm
 \cite{Sandvik:1999_honey_bilayer_QMC,Syljuasen:2002_honey_bilayer_QMC}.
 The value $\delta_{c}^{>}=1.645(1)$ is thereby obtained.  The
 independent result $\delta_{c}^{>}=1.66(1)$ has also been obtained
 \cite{Oitmaa:2012_honey_bilayer} using an Ising series expansion
 method around the N\'{e}el state out to 14th-order for $M$, and a
 dimer series expansion about the IDVBC state out to 10th-order for
 $\Delta$.  Our own best estimate, $\delta_{c}^{>} \approx 1.70(5)$,
 is in excellent agreement with these results.  We also remind the
 reader that our results for the position of the phase boundary at
 which N\'{e}el order melts are probably {\it least} accurate, in the
 entire $\kappa\delta$ plane, precisely along the $\kappa=0$ axis, for
 reasons we have given.

 It is interesting to compare these results with those obtained from
 different versions of mean-field theory (MFT).  One rather elegant
 such approach is the bond-operator formalism \cite{Sachdev:1990}, in
 which the spin operators are represented in a basis comprising
 singlet and triplet states on the interlayer NN bonds.  As we have
 noted, in the limit when the intralayer couplings vanish, the GS
 phase is simply the IDVBC state of localized singlets on these
 interlayer NN bonds $(J_{1}^{\perp})$.  At the mean-field level the
 IDVBC state is simply a uniform condensate of the singlet bosons, and
 the spin operators are then described in terms of the triplet
 excitations.  If we then rewrite the Hamiltonian for the
 $J_{1}$--$J_{1}^{\perp}$ model in terms of these triplet operators,
 and drop the corresponding quartic terms, which is tantamount to
 excluding triplet-triplet interactions, we arrive at the so-called
 singlet-triplet MFT.  It yields the value $\delta_{c}^{>}(0)=1.312$
 \cite{Ganesh:2011_honey_bilayer_PRB84}, which is somewhat below the
 presumably exact QMC value $\delta_{c}^{>}(0)=1.645(1)$.

 By contrast, Schwinger-boson MFT (SBMFT)
 \cite{Zhang:2014_honey_bilayer,Arlego:2014_honey_bilayer} yields the
 result $\delta_{c}^{>}(0) \approx 3.4$, which is now considerably
 {\it greater} than the QMC value.  It is interesting to note that the
 SBMFT results of Zhang {\it et al}. \cite{Zhang:2014_honey_bilayer}
 were also augmented by the same authors by spin-gap calculations
 using both a low-order (i.e., up to fourth-order) series expansion
 (SE) about the IDVBC limit and an exact diagonalization (ED) of a
 small (24-site) cluster.  Whereas the fourth-order SE calculation for
 the $\kappa=0$ case showed a tendency for the spin gap to close for
 values $J_{1} \gtrsim 0.62$ for
 $J_{1}^{\perp}=1$ (i.e., equivalent to
 $\delta > \delta_{c}^{>}(0) \approx 1.61$), the ED calculation shows
 no obvious tendency for the gap to close.  This is presumably due to
 very strong finite-size effects.  In any case, whereas the main SBMFT
 findings show a range of results for $E/N$, $M$, and $\Delta$ that
 are qualitatively similar to ours, the limited SE results
 \cite{Zhang:2014_honey_bilayer} are in closer quantitative agreement
 with ours, just like the more extensive SE calculations of Oitmaa and
 Singh \cite{Oitmaa:2012_honey_bilayer}.

 The SBMFT approach has also been used
 \cite{Zhang:2014_honey_bilayer,Arlego:2014_honey_bilayer} to give an
 estimate of the entire N\'{e}el phase boundary in the $\kappa\delta$
 phase for our spin-$\frac{1}{2}$ $J_{1}$--$J_{2}$--$J_{1}^{\perp}$
 model on the honeycomb bilayer lattice.  It also finds a reentrant
 behavior for small values of the interlayer coupling constant
 $\delta$.  However, the reentrant region in the parameter $\kappa$ is
 appreciably larger than that found by our CCM analysis.  Thus, the
 SBMFT values are $\kappa_{c}(0)=0.2075$ and $\kappa^{>}=0.289$,
 compared to our CCM estimates $\kappa_{c}(0)\approx 0.19(1)$ and
 $\kappa^{>} \approx 0.215(2)$.  Given that a higher-order CCM
 calculation is likely to be much more quantitatively accurate than
 any single-shot MFT approach, and given the large discrepancy of the
 SBMFT value for $\delta_{c}^{>}(0)$ from the essentially exact QMC
 value (which itself also agrees well with our CCM value), it seems
 likely that the larger SBMFT value for $\kappa^{>}$ is again an
 artefact of the MFT approach.  Nevertheless, it is gratifying that
 the overall shape of the N\'{e}el phase boundary so obtained in the
 SBMFT approach for the model in the $\kappa\delta$ phase agrees
 rather well with our CCM result in Fig.\ \ref{phase_diag}.

 In this paper we have considered the stability of the N\'{e}el phase
 alone.  One might imagine that in the large-$\delta$ region [say,
 $\delta > \delta_{c}^{>}(\kappa)$] there could also exist, for
 example, partially disordered phases that are mixtures of interlayer
 spin-singlet dimers and AFM ordering on specific sublattices.  Such
 states clearly cannot be ruled out by our present results.  To study
 them within the CCM framework would require the use of model states
 that are more complicated than the broad class of independent-spin
 product states with perfect magnetic LRO, the use of which has been
 discussed in Sec.\ \ref{ccm_sec}.  However, non-classical VBC
 ordering can also be directly considered within the CCM framework by
 employing valence-bond model states that are, for example, direct
 products of independent two-spin (dimer) or $n$-spin (plaquette)
 singlets \cite{Xian:1994_ccm_1Dchain}.  It is also interesting to
 note that it has been shown that dimer and plaquette VBC states for
 quantum magnets may actually even be formed via the usual CCM with
 independent-spin product model states \cite{Farnell:2009_review}.
 However, the use of either of these techniques or appropriate
 extensions of them would take us far outside the scope of the
 present work.

\section*{ACKNOWLEDGMENTS}
We thank the University of Minnesota Supercomputing Institute for the
grant of supercomputing facilities, on which the work reported here
was performed.  We also thank D. J. J. Farnell for his assistance in
extending the CCCM code \cite{ccm_code} appropriately.  One of us (RFB) gratefully
acknowledges the Leverhulme Trust (United Kingdom) for the award of an
Emeritus Fellowship (EM-2015-007).

\bibliographystyle{apsrev4-1}
\bibliography{bib_general}

\end{document}